\documentclass[prb,aps,superscriptaddress, twocolumn]{revtex4-2}
\usepackage{graphicx} 

\usepackage[T1]{fontenc}
\usepackage{times}
\usepackage{xcolor}
\usepackage[colorlinks,allcolors=blue]{hyperref}
\usepackage{graphicx}
\usepackage{bm}
\usepackage{mathrsfs}
\usepackage{dcolumn}
\usepackage{dsfont}
\usepackage{amsmath}
\usepackage{amsfonts}
\usepackage{amssymb}
\usepackage{mathtools}
\usepackage{braket}
\usepackage{tikz}
\usepackage{caption}
\usepackage{subcaption}
\usepackage{cancel}

\renewcommand{\epsilon}{{\varepsilon}}
\renewcommand{\vec}[1]{{\ensuremath{\bm{\mathrm{#1}}}}}

\renewcommand{\exp}[1]{\ensuremath{{\mathrm{e}^{#1}}}}
\newcommand{\rexp}{\operatorname{exp}}
\newcommand{\ii}{{\ensuremath{\mathrm{i}}}}
\newcommand{\rmd}{{\ensuremath{\mathrm{d}}}}
\newcommand{\kB}{{\ensuremath{k_{\mathrm{B}}}}}
\newcommand{\muB}{{\ensuremath{\mu_{\mathrm{B}}}}}
\renewcommand{\Im}{{\text{Im}}}
\renewcommand{\Re}{{\text{Re}}}

\newcommand{\will}[1]{\textcolor{red}{[Will: #1]}}
\newcommand{\classical}[1]{\vec{#1}^\text{cl}}
\newcommand{\quantum}[1]{\vec{#1}^\text{q}}

\newcommand{\soconst}{\lambda_{so}}

\newcommand{\fermi}[1]{N_\text{F}\left(#1\right)}
\newcommand{\angstrom}{\text{\normalfont\AA}}

\DeclareMathOperator{\tr}{Tr}

\begin{document}

\title{Microscopic theory of an atomic spin diode}

\author{William J. Huddie}
\affiliation{Institute for Theoretical Physics, Utrecht University, Princetonplein 5 3584CE Utrecht, The Netherlands}
\affiliation{Department of Applied Physics, Eindhoven University of Technology, P.O. Box 513, 5600MB, Eindhoven, The Netherlands}
\author{Rembert A. Duine}
\affiliation{Institute for Theoretical Physics, Utrecht University, Princetonplein 5 3584CE Utrecht, The Netherlands}
\affiliation{Department of Applied Physics, Eindhoven University of Technology, P.O. Box 513, 5600MB, Eindhoven, The Netherlands}

\date{\today}
\begin{abstract}
    We present a microscopic theory of an atomic spin diode. Our proposed system consists of two magnetic adatoms deposited on the surface of a two-dimensional electron gas with Rashba spin-orbit coupling. A local \textit{s-d} type coupling between the local spins and the spins of the electrons induces a non-local Ruderman-Kittel-Kazuya-Yoshida type interaction and a Dzyalonshinskii-Moriya interaction, in addition to dissipative interactions, between the spins. We derive the effective action for the spins using the Keldysh formalism. From the effective action, we also derive equations of motion for the spins which are shown to be of Landau-Lifshitz-Gilbert (LLG) type, and give expressions for the effective field and Gilbert damping which appear in this equation. From our microscopic theory, we find that for an in-plane magnetic field perpendicular to the vector connecting the two atoms, the magnitude of the field and the distance between the atoms can always be tuned to engender perfectly diodic coupling. Our findings may pave the way to experimental realisation of atomic spin diodes.
\end{abstract}

\maketitle

\section*{Introduction} The field of spintronics is concerned with the design and engineering of new devices which are based on magnetic materials, with the aim that these will eventually replace conventional electronics \cite{Barman_MagnonicsRoadmap2021,Flebus_MagnonicsRoadmap2024, Lenk_BuildingBlocksMagnonics}. 
Such electronic (CMOS--based) devices rely on the transistor, which only allows transport of charge across it when a gate voltage is applied. The device which underlies the operation of the transistor is called a pn junction, or semiconducting diode, which allows the flow of electric current across it in one direction, but strongly suppresses flow in the opposite direction -- this is an example of a non-reciprocal phenomenon \cite{Shockley_TheoryPNJunctions,Sze_SemiconductorDevices}. \\
In spintronics, however, one is interested in the transport of information using the spin of the electron as opposed to its charge. Such devices may transport information either using spin-polarised electric currents, or using magnons, the quanta of spin waves, which have the advantage that the propagation of such quasiparticles generates much less heat than is lost to Joule heating in ordinary electronics \cite{Chumak_MagnonSpintronics}. One may imagine, for example, a so--called magnon transistor, allowing the transmission of magnons across it only under certain conditions. The design of such a device would also necessarily have to include some diode component, facilitating this non-reciprocal behaviour. Non-reciprocal spin waves in certain systems have been known to exist since the 1960s \cite{DamonEshbach_MagnetostaticModes}, while recent experimental and theoretical works have demonstrated a variety of ways in which one may construct such a magnon transistor \cite{Chumak_MagnonTransistor, Lan_SpinWaveDiode, Yuan_PhysRevB_107_024418_2023, Zou_PhysRevLett_132_036701_2024}.  \\
Ref. \cite{Yuan_PhysRevB_107_024418_2023} demonstrates, phenomenologically at least, the existence of a perfectly diodic (unidirectional) coupling in an archetypal synthetic antiferromagnet (SAF) structure, in which the two ferromagnetic layers are coupled via a coherent magnetic interaction (e.g. the Dzyaloshinskii-Moriya interaction (DMI)), and an incoherent interaction (e.g. spin pumping across the spacer layer). The authors show that, under certain assumptions, the magnetisation dynamics of each magnetic layer are qualitatively different; that is, when one layer is driven with a microwave field, the other layer undergoes its dynamics in response, but not vice-versa.
This phenomenon has subsequently also been corroborated using the Lindblad master equation formalism \cite{Zou_PhysRevLett_132_036701_2024}. The key ingredients for the perfectly diodic behaviour emphasised in Refs. \cite{Yuan_PhysRevB_107_024418_2023, Zou_PhysRevLett_132_036701_2024} are DMI  and dissipative interactions, which precisely cancel one another for a specific field. This is an implementation in a magnetic system of the ideas put forward in Ref. \cite{PhysRevX.5.021025} for coupling photonic cavities.  \\
In this work, we consider a rather different system from that considered in \cite{Yuan_PhysRevB_107_024418_2023, Zou_PhysRevLett_132_036701_2024}, but which still possesses all the key ingredients to realise non-reciprocal spin dynamics. The setup, which is depicted in Fig. \ref{fig:syssetup} we consider a system of two atomic spins (magnetic adatoms) deposited on a two-dimensional electron gas (2DEG) substrate. Rashba-type spin-orbit coupling is present, and we assume a \textit{s-d} type coupling between the localised spins and the spins of the electron gas. \\ 
Several groups have recently demonstrated the experimental feasibility of constructing systems similar to the one we consider here by use of scanning-tunnelling microscopy (STM) techniques \cite{khajetoorians2012atom, khajetoorians2019creating, Choi_RevModPhys_91_041001_2019}. Not only are such one-dimensional atomic spin chains relatively straightforward to construct, it is also possible to measure their spin-wave dispersion by similar techniques \cite{Spinelli_NatMater_13_782_2014}.  \\ 
Using a Keldysh approach to determine the effective dynamics of the atomic spins, we find that these dynamics are of Landau-Lifshitz-Gilbert (LLG) type. The effective equations of motion contain non-local coherent and dissipative interactions, as well as local Gilbert damping. Moreover, examining the magnon susceptibility, we find that for an in-plane magnetic field perpendicular to the vector connecting the two atoms, the magnitude of the field and the distance between the atoms can always be tuned to engender unidirectional coupling. These findings may prove beneficial in the future experimental realisation of atomic spin diodes. \\
The outline of this work is as follows. In Se. \ref{sec:msystem}, we give a description of the model system and Hamiltonian, and summarise our key results. \\
In Sec. \ref{sec:effaction}, we develop the action of this system along the Keldysh contour. Using perturbation theory in the \textit{s-d} coupling constant, we integrate out the electronic bath and derive an effective action on the spins. \\
In Sec. \ref{sec:linres}, we expand the electronic response function to first order in frequency, allowing us to derive exact expressions for the former. \\
In Sec. \ref{sec:llg}, we relate the real part of the retarded and advanced components of this response function to the non-local  Ruderman-Kittel-Kasuya-Yosida (RKKY)--type coupling between the spins (including an antisymmetric component, which is identified as the DMI), while the imaginary part is related to the Gilbert damping. We show that the equations of motion of the spins are exactly in LLG form, and study the linearised forms of these equations. 
We end with a short conclusion and discussion of our results, and provide some outlook on possible continuations of our work.
\section{Model and summary of results} \label{sec:msystem}
\begin{figure}
    \begin{center}
    \includegraphics[width=1.05\linewidth]{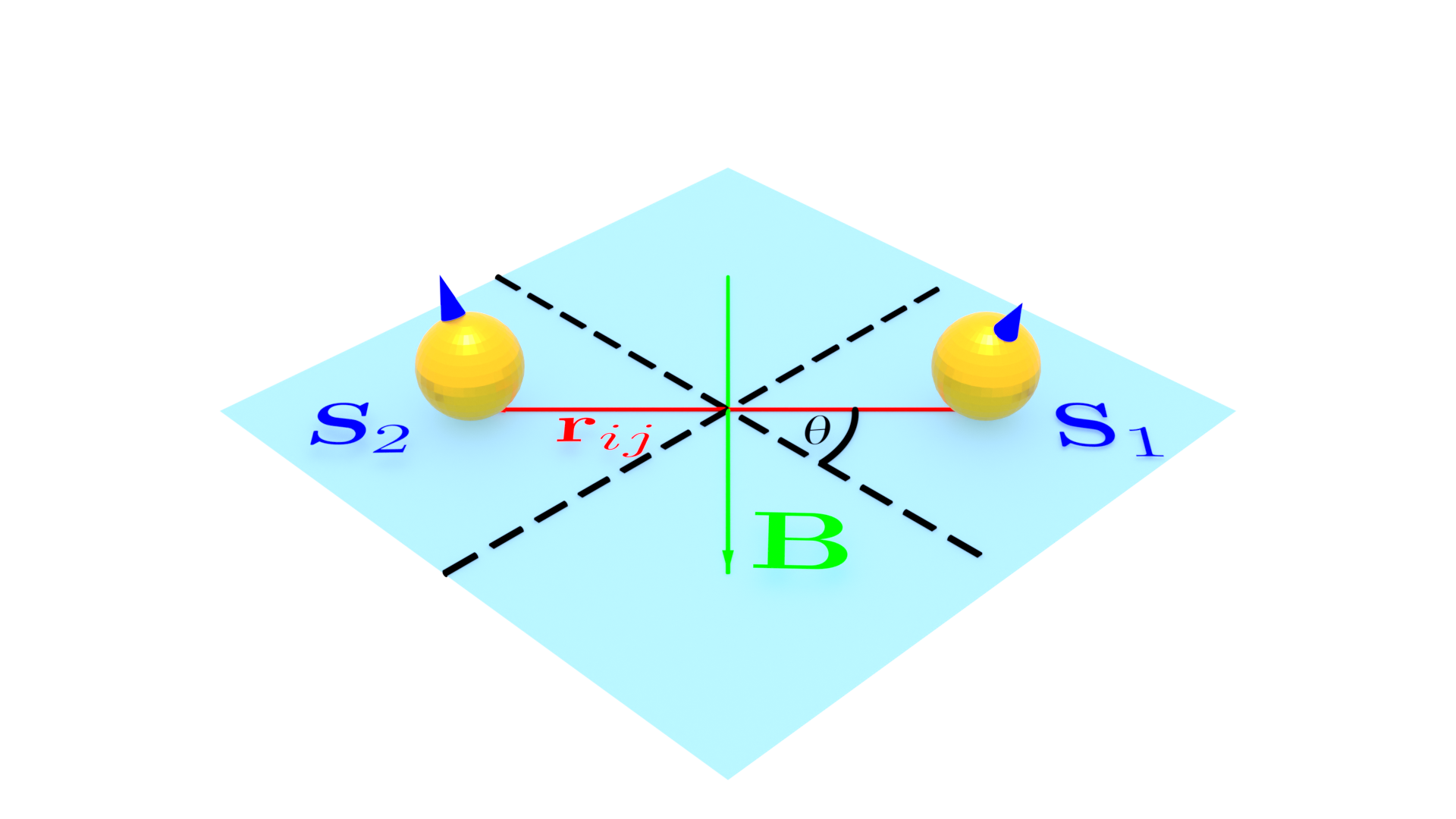}
    \end{center}
    \caption{Two atomic spins are deposited on the surface of a two-dimensional electron gas (2DEG). Localised \textit{s-d} type coupling to an electron bath induces coherent and dissipative interactions between the two spins}
    \label{fig:syssetup}
\end{figure}
In this section, we introduce the model that is considered in this work. We give the Hamiltonian of this model system, and give a brief description of the methodology which is used to obtain the semiclassical equations of motion. Readers who are mainly interested in the primary results of our work may want to skip Sections \ref{sec:effaction} and \ref{sec:linres}, and to this end we quote here the equations of motion and the conditions we derive on the parameters which appear in the latter such that the magnon transmission between the two spins is unidirectional. \\
The setup we consider is depicted in Fig. \ref{fig:syssetup}.
The Hamiltonian of this model system is given by 
\begin{equation}
    \hat{\mathcal{H}} = \hat{\mathcal{H}}_\text{S} + \hat{\mathcal{H}}_\text{B} + \hat{\mathcal{H}}_{\text{SB}},
\end{equation}
which we have split into system, bath, and system-bath interaction parts, in the context of an open quantum system \cite{BreuerPetruccione_OpenQuantumSystems}. This Hamiltonian consists of three parts. The first term, $\hat{\mathcal{H}}_\text{S}$ is the spin Hamiltonian of the localised spins; we write this as
\begin{equation}
    \hat{\mathcal{H}}_\text{S} = -\sum_i\vec{B}_i\cdot\hat{\vec{S}}_i,
\end{equation}
where $\vec{S}_i$ are the localised spins at sites $1$ and $2$. The sum is taken over spin sites $i=1,2$, and we have absorbed the units of $g\muB$ into the definition of $\vec{B}_i$. \\
The second term $\hat{\mathcal{H}}_\text{B}$ is the Hamiltonian of a two-dimensional electron gas with Rashba spin-orbit coupling; it is given by
\begin{equation}
    \hat{\mathcal{H}}_\text{B} = \int\rmd^2\vec{r} \ \hat{\psi}^\dagger(\vec{r})\left(\frac{\hat{\vec{p}}^2}{2m} + \soconst(\hat{\vec{z}}\times\hat{\vec{p}})\cdot\vec{\sigma}\right)\hat{\psi}(\vec{r}),
\end{equation}
where $\hat{\psi}^\dagger, \hat{\psi}$ are Weyl spinors in the standard $\sigma^z$ eigenbasis $\{\ket{\uparrow}, \ket{\downarrow}\}$, $\hat{\vec{p}} = -\ii\vec{\nabla}$ is the momentum operator with $m$ the effective electron mass, $\hat{\vec{z}}$ is the unit vector in the direction perpendicular to the plane of the 2DEG, and $\vec{\sigma} = (\sigma^x, \sigma^y, \sigma^z)^T$ is a vector of Pauli matrices. We shall hereafter refer to $\hat{\mathcal{H}}_\text{B}$ as the bath Hamiltonian, and to the 2DEG itself as the bath.\\
The last term, $\hat{\mathcal{H}}_\text{SB}$, is the system-bath interaction. This is assumed to be of \textit{s-d} type and is of the general form \cite{Yosida_TheoryMagnetism}
\begin{equation}
    \hat{\mathcal{H}}_\text{SB} = -\int\rmd^2\vec{r} \ J_{sd}(\vec{r})\vec{S}(\vec{r})\cdot\vec{s}(\vec{r}), \label{eq:hamsd}
\end{equation}
where $J_{sd}(\vec{r})$ is the strength of the \textit{s-d} coupling, $\vec{S}(\vec{r})$ is a localised spin at position $\vec{r}$, with $\vec{s}(\vec{r})$ the electronic spin density at that same point. \\
For simplicity, we assume a contact ($\delta$-function type) interaction between localised spins and itinerant electrons, i.e. we write $J_{sd}(\vec{r}) = J_{sd}\sum_i \delta(\vec{r} - \vec{r}_i)$, which reduces Eq. (\ref{eq:hamsd}) to
\begin{equation}
    \hat{\mathcal{H}}_\text{SB} = -J_{sd}\sum_i\sum_\mu \hat{S}_i^\mu \hat{s}^\mu_i.
\end{equation}
Note that the dimensions of $J_{sd}$ are $[\text{energy}]\times[\text{area}]$. \\
After performing the Fourier transform, we find
\begin{align}
    \hat{\mathcal{H}}_\text{S} &= -\sum_{i=1,2}\vec{B}_i\cdot\vec{S}_i, \label{eq:hamsys}\\
    \hat{\mathcal{H}}_\text{B} &= \sum_{\vec{k}s}\epsilon_{\vec{k}s}\hat{c}^\dagger_{\vec{k}s}\hat{c}_{\vec{k}s}, \label{eq:hambath} \\
    \hat{\mathcal{H}}_{\text{SB}} &= -J_{sd}\sum_{i}\sum_\mu \hat{S}_i^\mu \hat{s}^\mu_i\label{eq:hamsysbath},
\end{align}
Eqs. (\ref{eq:hambath}-\ref{eq:hamsysbath}) are written in the Rashba basis $\{\hat{c}^\dagger_{\vec{k}s}, \hat{c}_{\vec{k}s}\}$, where $\vec{k} = (k_x, k_y)^T, \ s = \pm 1$ is the helicity index and ($\hbar = 1$)
\begin{align}
    \epsilon_{\vec{k}s} &= \frac{k^2}{2m} + s\soconst k = \epsilon_k + s\soconst k  \\
\end{align}
is the dispersion relation of electrons with momentum $\vec{k}$ in the Rashba band $s = \{+, -\}$, whereas the creation and annihilation operators in this basis are related to those in the spin-z eigenbasis $\sigma = \{\uparrow, \downarrow\}$ via
\begin{align}
    \begin{pmatrix}
        \hat{c}_{\vec{k}+} \\
        \hat{c}_{\vec{k}-} \\
    \end{pmatrix} &= U^\dagger_\vec{k}\begin{pmatrix}
        \hat{c}_{\vec{k}\uparrow} \\
        \hat{c}_{\vec{k}\downarrow}
    \end{pmatrix} = \frac{1}{\sqrt{2}}\begin{pmatrix}
        \ii\exp{\ii\phi_k} & 1 \\
         -\ii\exp{\ii\phi_k} & 1
    \end{pmatrix}\begin{pmatrix}
        \hat{c}_{\vec{k}\uparrow} \\
        \hat{c}_{\vec{k}\downarrow}
    \end{pmatrix} \label{eq:rashbabasis}
\end{align}
and we defined for conciseness
\begin{align}
    k &= \lvert\vec{k}\rvert = \sqrt{k_x^2 + k_y^2}; \nonumber \\
    \phi_k &= \arctan\left(\frac{k_y}{k_x}\right), \nonumber
\end{align}
In the basis (\ref{eq:rashbabasis}), the components of the electronic spin density $\hat{s}^\mu_i$ are given by
\begin{align}
    \hat{s}^\mu_i &= \frac{1}{A}\sum_{\vec{k}\vec{k'},ss'}\exp{-\ii(\vec{k}-\vec{k'})\cdot\vec{r}_i}\Braket{s,\vec{k}|\sigma^\mu|s',\vec{k'}}\hat{c}^\dagger_{\vec{k}s}\hat{c}_{\vec{k'}s'}.
\end{align}
Using the Keldysh path integral formalism \cite{Kamenev_FieldTheoryNonEq}, we develop the action of this model system. We integrate out the electron bath using second-order perturbation theory in the \textit{s-d} coupling constant $J_{sd}$, thereby obtaining an effective action on the spins which is written in terms of the spin-density spin-density response function of the bath. \\
In the frequency domain, we expand the electronic response function to first order in frequency and show that the effective action yields the following equations of motion: 
\begin{equation}
    \dot{\vec{S}}_k(t) = \vec{S}_k(t)\times\left[\vec{B}^\text{ext} + \sum_j(\vec{\Gamma}_{ij}*\vec{S}_j)(t) + \vec{\xi}_j(t)\right] \label{eq:llgnonsimp},
\end{equation}
where the friction kernel $\vec{\Gamma}_{ij}$ is obtained from the microscopic properties of the bath; in the low-frequency regime, Eq. (\ref{eq:llgnonsimp}) simplifies to
\begin{equation}
    \dot{\vec{S}}_k(t) = \vec{S}_k(t)\times\left[\vec{B}^\text{ext} + \sum_j\vec{K}_{kj}\vec{S}_j(t) - \vec{\alpha}_{kj}\dot{\vec{S}}_j(t) + \vec{\xi}_j(t)\right] \label{eq:llg},
\end{equation}
which is in the form of a generalised LLG equation.
The exchange tensor $\vec{K}_{ij}$ and the Gilbert damping tensor $\vec{\alpha}_{ij}$ contain both local ($i = j$) and non-local ($i \neq j $) terms. The relation of these terms to the microscopic properties of the bath is done in Sec. \ref{sec:effaction}, while explicit expressions are given in Sec. \ref{sec:linres}. \\
Taking Eq. (\ref{eq:llg}), we study the magnon susceptibility, and derive a condition on (a) the interatomic spacing $R = \lvert\vec{r}_{12}\rvert = \lvert\vec{r}_{21}\rvert$ and (b) the applied magnetic field $B_0 = \lvert\vec{B}^\text{ext}\rvert$ such that the magnon transmission between spin $1$ and spin $2$, or vice versa, is entirely unidirectional. Details of this calculation, and of the specific conditions we derive, are given in Sec. \ref{sec:llg}.  
\section{Equations of motion} \label{sec:effaction}
In this section, we derive the equations of motion for the semiclassical spin dynamics of the system. We begin with a brief overview of the Keldysh path integral formalism, which is a widely-used method for applying the techniques of quantum field theory to non-equilibrium systems, and finds diverse applications in a variety of areas of physics. We then develop the combined action of the system and bath, integrate out the latter, and derive the equations of motion from the resulting effective action on the former. We find that these equations of motion are in the form of a generalised Landau-Lifshitz-Gilbert equation.
\subsection{Keldysh path integral}
\begin{figure}[h!]
    \begin{center}
    \begin{tikzpicture}
        \draw[->] (4,0) -- (0,0) node[midway, below] {$-$};
        \node[left] at (0,0) {$t=-\infty$};

        \draw[->] (0,-1) -- (4,-1) node[midway, above] {$+$};
        \node[left] at (0,-1) {$t=-\infty$};

        \draw (4,-1) to[out=0, in=0] (4,0);

        \node[right] at (4.5, -0.5) {$t=+\infty$};

        \node[below] at (2.5,-1.05) {$t'$};
        \node at (4.2,-1.2) {$\mathcal{C}$};
    \end{tikzpicture}
    \end{center}
    \caption{Schematic representation of the time evolution along the Keldysh contour $\mathcal{C}$.}
    \label{fig:keldyshcont}
\end{figure}
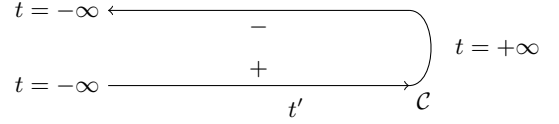
To derive an expression for the real-time action, one considers the coherent-state path integral along the Keldysh time contour, depicted schematically in Fig. \ref{fig:keldyshcont}. The system is initially in thermal equilibrium, and an interaction between the system of interest and some environment is adiabatically switched on from $t=-\infty$, reaching its full strength at $t = +\infty$. The system then adiabatically returns to thermal equilibrium at $t=-\infty$ \cite{Kamenev_FieldTheoryNonEq}. For similar techniques which are applied to study the dynamics of spin systems coupled to an environment, one may consult Refs. \cite{Duine_PhysRevB_75_214420_2007, Nunez_PhysRevB_77_054401_2008, Verstraten_PhysRevRes_5_033128_2023, Quarenta_PhysRevLett_133_136701_2024}. \\
We develop the coherent path integral over fermionic (Grassman) fields $\{\psi^*_{\vec{k}\alpha}, \psi_{\vec{k}\alpha}\}$ and coherent spin states $\ket{g_i}$ along the contour $\mathcal{C}$, and we find that the partition function is given by:
\begin{align}
    Z[\vec{S}_1, \vec{S}_2, t] &= \int \mathcal{D}[g_1] \mathcal{D} [g_2]\int\mathcal{D}[\psi^*]\mathcal{D}[\psi] \nonumber \\ 
    &\times\rexp\left(\ii\mathcal{S}[\vec{S}_1, \vec{S}_2, \psi^*, \psi]\right) \label{eq:partfunc},
\end{align}
where $\vec{S}_i(t) = \braket{g_i|\hat{\vec{S}}_i(t)|g_i}$ is the expectation value of the spin operator $\hat{\vec{S}}_i$ in the spin-coherent state basis which is defined by the Euler angles $(\psi_i, \theta_i, \phi_i)$ as \cite{AltlandSimons_CondMatFieldTheory}
\begin{align}
    \ket{g_i} &\equiv \exp{-\ii\phi_iS^z_i}\exp{-\ii\theta_iS^y_i}\exp{-\ii\psi_iS^z_i}\ket{S;S} \nonumber \\
    &= \exp{-\ii\psi_i S}\exp{-\ii\phi_iS^z_i}\exp{-\ii\theta_iS^y_i}\ket{S;S} \label{eq:spincostat},
\end{align}
where $\ket{S;S}$ is the state with maximal spin, with $S$ the spin quantum number. Note that the Euler angle $\psi_i$ gives an overall phase and hence a gauge freedom to the spin coherent state. The choice of gauge is important, as certain choices of gauge will cause certain paths taken by the spins to vanish pathologically \cite{Verstraten_PhysRevRes_5_033128_2023}. We discuss this gauge freedom further in Section \ref{subsec:keldrot}.\\
The action $\mathcal{S}[\vec{S}_1, \vec{S}_2, \psi^*, \psi]$ which weights the path integral in Eq. (\ref{eq:partfunc}) is written as the sum of three parts:
\begin{align}
    \mathcal{S}[\vec{S}_1, \vec{S}_2, \psi^*, \psi] &= \mathcal{S}_\text{S}[g_1, g_2] \nonumber \\ 
    &+ \mathcal{S}_\text{B}[\psi^*, \psi] \nonumber \\
    &+ \mathcal{S}_\text{SB}[g_1, g_2, \psi^*, \psi],
\end{align}
where
\begin{align}
    \mathcal{S}_\text{S} &= \sum_{i}\oint_{\mathcal{C}}\rmd t \ \left[\Braket{g_i|\ii\partial_t|g_i} - \vec{B}_i\cdot\vec{S}_i\right], \label{eq:sysaction} \\
    \mathcal{S}_\text{B} &= \oint_\mathcal{C}\rmd t \ \sum_{\vec{k}\alpha}\psi^*_{\vec{k}\alpha}(t)\left(\ii\partial_t -\left( \epsilon_{\vec{k}\alpha}-\mu\right)\right)\psi_{\vec{k}\alpha}(t),\label{eq:bathaction}
\end{align}
\begin{equation}
    \mathcal{S}_\text{SB} = J_{sd}\oint_\mathcal{C}\rmd t\sum_{i}\sum_{\mu}S^\mu_i(t)s^\mu_i(t) \label{eq:sysbathaction},
\end{equation}
and the time-dependent electronic spin density is given by
\begin{align}
    s^\mu_i(t) &= \frac{1}{A}\sum_{\vec{k}\vec{k'},ss'}\exp{-\ii(\vec{k}-\vec{k'})\cdot\vec{r}_i}\Braket{s,\vec{k}|\sigma^\mu|s',\vec{k'}}\psi^*_{\vec{k}s}(t)\psi_{\vec{k}s'}(t)
\end{align}
To obtain an effective action on the spins, we integrate out the fermionic fields. In order to do this, we assume that the \textit{s-d} coupling constant $J_{sd}$ can be treated perturbatively, such that we can expand the term $\exp{\ii\mathcal{S}_\text{SB}}$ in powers of this quantity; this is a good approximation provided the spin-flip scattering amplitude density is small \cite{kondo1964resistance, Anderson_PhysRevB_1_4464_1970}, related to the condition that the system is measured at a temperature much larger than the Kondo temperature \cite{Turco_PhysRevRes_6_L022061_2024,Zhang_NatCommun_4_2110_2013, Madhavan_PhysRevB_64_165412_2001, Hiraoka_NatCommun_8_16012_2017, Wahl_PhysRevLett_98_056601_2007, wan2023evidence} -- i.e., we must be in a temperature regime where the RKKY interaction is dominant over the Kondo effect. \\
The validity of this expansion is not controlled by the parameter $J_{sd}$ itself, but rather by the quantity $J_{sd}\rho$, where $\rho$ is the electronic density of states at the Fermi level; each power of the electronic spin density brings with it a factor of the (constant) density of states, and $J_{sd}\rho \ll 1$ is typically a very good approximation. \\
Expanding to second order in $J_{sd}$ and performing the path integral over the fermionic fields, we find that the first-order term vanishes (physically, the Rashba 2DEG has no spontaneous spin-polarisation), whereas the zeroth and second-order terms can be re-exponentiated to obtain an effective action on the spins of the form
\begin{equation}
    \ii\mathcal{S}_\text{eff}[g_1, g_2] = \ii\mathcal{S}_0[g_1, g_2] + \ii\mathcal{S}_\text{int}[g_1, g_2],
\end{equation}
where
\begin{align}
    \ii\mathcal{S}_\text{int} &= \ii\oint_C\rmd t\oint_C\rmd t'\sum_{ij}\sum_{\mu\nu} \nonumber \\
    &\times S^\mu_i(t)\vec{\Pi}^{\mu\nu}_{ij}(t,t')S^\nu_j(t') \label{eq:sint}.
\end{align}
Here, we defined the electronic spin-density spin-density response function $\vec{\Pi}^{\mu\nu}_{ij}(t,t')$ as
\begin{align}
    \vec{\Pi}^{\mu\nu}_{ij}(t,t') &= \frac{\ii J_{sd}^2}{2}\Braket{s^\mu_i(t) s^\nu_j(t')}_0 \nonumber \\
    &= \frac{\ii J_{sd}^2}{2}\int\frac{\rmd^2\vec{k}}{(2\pi)^2}\int\frac{\rmd^2\vec{k'}}{(2\pi)^2} \exp{-\ii(\vec{k}-\vec{k'})\cdot\vec{r}_{ij}}\nonumber \\
    &\times \sum_{ss'}\mathcal{M}^{\mu\nu}_{\vec{k}\vec{k'},ss'} G_s(\vec{k},t,t')G_{s'}(\vec{k'},t',t)\label{eq:responsefunc},
\end{align}
where, $G_s$ is the electronic Green's function on the Keldysh contour, given by
\begin{equation}
    G_{ss'}(\vec{k},\vec{k'}; t,t') = \Braket{\psi_s(\vec{k}, t)\psi_{s'}^*(\vec{k'},t')}_0\delta_{ss'}\delta_{\vec{k}\vec{k'}} \label{eq:gfunc},
\end{equation}
and we defined the matrix element $\mathcal{M}^{\mu\nu}_{\vec{k}\vec{k'},ss'}$ as 
\begin{equation*}
    \mathcal{M}^{\mu\nu}_{\vec{k}\vec{k'},ss'} = \Braket{s,\vec{k}|\sigma^\mu|s',\vec{k'}}\Braket{s',\vec{k'}|\sigma^\nu|s,\vec{k}}
\end{equation*}
The subscript $0$ which appears in Eqs. (\ref{eq:responsefunc}--\ref{eq:gfunc}) implies that the expectation values are taken with respect to the initial equilibrium state of the bath, i.e. without any system-bath interactions, such that time-reversal and inversion symmetry are preserved. We also assumed that the bath is in the thermodynamic limit, such that the momentum space sums can be replaced by integrals. \\
Having derived a formal expression for the effective action, the next step is to extract from this measurable real-time quantities. The prescription for converting quantities on the Keldysh contour to measurable observables is called the Keldysh rotation \cite{Kamenev_FieldTheoryNonEq}, which is the subject of the following section.
\subsection{Keldysh rotation} \label{subsec:keldrot}
We begin with some basic rules for calculus of observables which live on the Keldysh contour. \\
Any observable $\mathcal{O}$ on the Keldysh contour can be decomposed into its greater and lesser parts via
\begin{equation}
    \mathcal{O}(t,t') = \Theta(t,t')\mathcal{O}^>(t,t') + \Theta(t',t)\mathcal{O}^<(t,t').
\end{equation}
The Keldysh rotation then lets us write down the retarded, advanced, and Keldysh components of any given observable, defined by
\begin{align} 
    \mathcal{O}^{R(A)}(t-t') &= \pm\Theta(\pm(t-t'))\left(\mathcal{O}^>(t-t') - \mathcal{O}^<(t-t')\right) \\
    \mathcal{O}^K(t-t') &= \mathcal{O}^>(t-t') + \mathcal{O}^<(t-t').
\end{align}
Next, we split the observable into its so-called classical and quantum parts, according to
\begin{equation}
\mathcal{O}^{\pm}(t) = \mathcal{O}^\text{cl}(t) \pm\frac{\mathcal{O}^\text{q}}{2}.
\end{equation}
These intuitively represent the average value of an observable and its fluctuating part, respectively. \\
We first focus on the system part of the action, Eq. (\ref{eq:sysaction}). Differentiation of Eq. (\ref{eq:spincostat}) gives us
\begin{equation}
    \Braket{g_i|\ii\partial_t|g_i} = S\left[\dot{\psi}_i + \dot{\phi}_i\cos\theta_i\right],
\end{equation}
where $S$ is the spin quantum number. The magnetic field term in Eq. (\ref{eq:sysaction}) becomes, if we assume $\vec{B}_i = (0,0,B^z_i)^T$,
\begin{equation}
    \Braket{g_i|\vec{B}_i\cdot\hat{\vec{S}}_i|g_i} = SB^z_i\cos\theta_i,
\end{equation}
from which we can see that $\mathcal{S}_\text{S}$ is
\begin{equation}
    \mathcal{S}_S = S\sum_i\oint_\mathcal{C}\rmd t \ \left[\dot{\psi}_i + \dot{\phi}_i\cos\theta_i + B^z_i\cos\theta_i\right] \label{eq:sysactionangles}
\end{equation}
Here is where the choice of gauge becomes important. We choose, following Refs. \cite{Verstraten_PhysRevRes_5_033128_2023} and \cite{Shnirman_PhysRevLett_114_176806_2015}, $\psi = \chi - \phi$, and $\dot{\chi}^\text{cl} = \dot{\phi}^\text{cl}(1 - \cos\theta^\text{cl})$, $\chi^\text{q} = \phi^\text{q}(1 - \cos\theta^\text{q})$. Under this choice of gauge, Eq. (\ref{eq:sysactionangles}) becomes, after Keldysh rotation,
\begin{widetext}
\begin{equation}
    \mathcal{S}_\text{S} = S\sum_i\int\rmd t \ \left[\phi^\text{q}_i\dot{\theta}^\text{cl}_i\sin\theta^\text{cl}_i\cos\frac{\theta^\text{q}_i}{2} + \left(\phi^\text{q}_i\frac{\dot{\theta}^\text{q}_i}{2}\cos\theta^\text{cl}_i - 2\dot{\phi}^\text{cl}_i\sin\theta^\text{cl}_i - 2B_i\sin\theta^\text{cl}_i\right)\sin\frac{\theta^\text{q}_i}{2}\right], \label{eq:sysactionqandcl}
\end{equation}
\end{widetext}
where henceforth, a time integral without the contour subscript denotes an integral over the real line $t \in \mathds{R}$. \\
We may make one further simplification to this action. Noting that our semiclassical Euler-Lagrange equations will be obtained by variations with respect to quantum components $\phi^\text{q}, \theta^\text{q}$, we may expand (\ref{eq:sysactionqandcl})  to first order in these components: 
\begin{equation}
    \mathcal{S}_\text{S} = S\int\rmd t \ \left[\phi^\text{q}_i\dot{\theta}^\text{cl}_i\sin\theta^\text{cl}_i - \theta^\text{q}_i\left(B_i +\dot{\phi}^\text{cl}_i\right)\sin\theta^\text{cl}_i\right].
\end{equation}
It is instructive to compute the classical and quantum components of $\Braket{g_i|\hat{\vec{S}}_i(t)|g_i} \equiv \vec{S}_i(t)$. To first order in quantum components, we find
\begin{align}
    \classical{S}_i &= S\begin{pmatrix}
        \sin\theta^\text{cl}_i\cos\phi^\text{cl}_i \\
        \sin\theta^\text{cl}_i\sin\phi^\text{cl}_i \\
        \cos\theta^\text{cl}_i
    \end{pmatrix}, \\
    \quantum{S}_i &= S\begin{pmatrix}
        \theta^\text{q}_i\cos\theta^\text{cl}_i\cos\phi^\text{cl}_i - \phi^\text{q}_i\sin\theta^\text{cl}_i\sin\phi^\text{cl}_i \\
        \theta^\text{q}_i\sin\theta^\text{cl}_i\cos\phi^\text{cl}_i + \phi^\text{q}_i\cos\theta^\text{cl}_i\sin\phi^\text{cl}_i \\
        -\theta^\text{q}_i\sin\theta^\text{cl}_i
    \end{pmatrix},
\end{align}
giving us the useful relation
\begin{equation} \label{eq:clqrelation}
        \quantum{S}_i = \theta^\text{q}_i\frac{\partial\classical{S}_i}{\partial\theta^\text{cl}_i} + \phi^\text{q}_i\frac{\partial\classical{S}_i}{\partial\phi^\text{cl}_i}.
\end{equation}
Performing the Keldysh rotation of $\mathcal{S}_\text{int}$ is no more difficult, though slightly more involved, as this term contains a double time integral along the Keldysh contour. One obtains:
\begin{align}
    \ii\mathcal{S}_\text{int} &= \ii\sum_{ij}\int\rmd t\int\rmd t' \ \left[\right.\quantum{S}_i(t)\cdot\left[\vec{\Pi}^R_{ij}(t-t')\classical{S}_j(t')\right] \nonumber \\
    &+ \ii\sum_{ij}\int\rmd t\int\rmd t' \ \classical{S}_j(t)\cdot\left[\vec{\Pi}^A_{ij}(t-t')\quantum{S}_i(t')\right] \nonumber \\
    &+\ii\sum_{ij}\int\rmd t\int\rmd t' \ \frac{1}{2}\quantum{S}_i(t)\vec{\Pi}^K_{ij}(t-t')\quantum{S}_j(t')\left.\right], \label{eq:intactionqandcl}
\end{align}
where the retarded, advanced, and Keldysh components of the response function are given by
\begin{align}
    \vec{\Pi}^{R(A)}(t-t') &= \pm\Theta(\pm(t-t'))\left(\vec{\Pi}^>(t-t') -\vec{\Pi}^<(t-t')\right); \label{eq:radefs} \\
    \vec{\Pi}^K(t-t') &= \vec{\Pi}^>(t-t') +\vec{\Pi}^<(t-t'). \label{eq:kelddefs} 
\end{align}
Having converted the contour integrals into real-time integrals, we are in a position to study the dynamics of the spins under the influence of the coupling to the electron bath. The derivation of these equations is the subject of the following section.
\subsection{Equations of motion}
The third term in (\ref{eq:intactionqandcl}) is quadratic in quantum components of the spins. In order to remove this term, we perform a Hubbard-Stratonovich transformation \cite{Hubbard_PhysRevLett_3_77_1959, Stoof_QuantumFluids}, which exactly cancels this term at the cost of introducing a pair of auxilliary fields $\vec{\xi}_1$ and $\vec{\xi}_2$. The partition function of the system becomes
\begin{align}
    Z[\vec{S}_1, \vec{S}_2, t] &= \int\mathcal{D}[\vec{\xi}_1]\int\mathcal{D}[\vec{\xi}_2]\int\mathcal{D}[g_1]\int\mathcal{D}[g_2] \nonumber \\
    &\times\rexp\left(\ii\mathcal{S}_\text{eff}\left[\hat{\vec{S}}_1, \hat{\vec{S}}_2, \vec{\xi}_1, \vec{\xi}_2\right]\right),
\end{align}
with the effective action
\begin{widetext}
    \begin{align}
        \ii\mathcal{S}_\text{eff} &= \ii S\sum_i \int\rmd t \ \left[\phi^\text{q}_i\dot{\theta}^\text{cl}_i\sin\theta^\text{cl}_i - \theta^\text{q}_i\left(B_i +\dot{\phi}^\text{cl}_i\right)\sin\theta^\text{cl}_i\right] -\frac{\ii}{2} \sum_{ij}\int\rmd t\ \vec{\xi}_i(t)\left(\left(\vec{\Pi}^K_{ij}\right)^{-1}*\vec{\xi}_j\right)(t) \nonumber \\
        &+\ii\sum_{ij}\int\rmd t \ \classical{S}_i(t)\cdot\left[\left(\vec{\Pi}^R_{ij}*\classical{S}_i\right)(t)\right] + \quantum{S}_j(t)\cdot\left[\left(\vec{\Pi}^A_{ij}*\classical{S}_i\right)(-t)\right] \nonumber \\
        &+ \ii\sum_{ij}\int\rmd t \ \quantum{S}_i(t)\cdot\vec{\xi}_j(t), \label{eq:effaction}
    \end{align}
\end{widetext}
where we rewrote the integral over $t'$ as a convolution. \\
One observes from Eq. (\ref{eq:effaction}) that i) the fluctuating magnetic fields have a noise spectrum defined by the Keldysh component of the magnetic field, namely
\begin{equation}
    \Braket{\vec{\xi}_i(t)\vec{\xi}_j(t')} = -\ii\vec{\Pi}^K_{ij}(t-t'),\label{eq:xicorrels}
\end{equation}
and ii) the Lagrangian of the system is given by
\begin{align}
    \mathcal{L}(t) &= S\sum_i\left[\phi^\text{q}_i\dot{\theta}^\text{cl}_i\sin\theta^\text{cl}_i - \theta^\text{q}_i\left(B_i +\dot{\phi}^\text{cl}_i\right)\sin\theta^\text{cl}_i\right] \nonumber \\
    &+ \sum_{ij} \quantum{S}_i(t)\cdot\left(\vec{\Pi}^R_{ij}*\classical{S}_i\right)(t) + \quantum{S}_j(t)\cdot\left(\vec{\Pi}^A_{ij}*\classical{S}_i\right)(-t) \nonumber \\
    &+ \sum_{ij}\quantum{S}_i(t)\cdot\vec{\xi}_j(t).
\end{align}
From the Lagrangian, we obtain the equations of motion by imposing the constraints
\begin{equation}
    \frac{\partial\mathcal{L}}{\partial\phi^\text{q}_k} = \frac{\partial\mathcal{L}}{\partial\theta^\text{q}_k} = 0 \quad k = 1,2.
\end{equation}
Note that any dependence on the time derivatives of $\theta^\text{q}$, $ \phi^\text{q}$ is removed by the expansion we made to first order in these terms. \\
We define the friction kernel:
\begin{equation*}
    \vec{\Gamma}_{kj}(t-t') = \vec{\Pi}^R_{kj}(t-t') + \vec{\Pi}^A_{jk}(t'-t).
\end{equation*}
If we make use of Eq. (\ref{eq:clqrelation}) and rearrange a bit, we find that these constraints give us the equations of motion for $\phi^\text{cl}_k, \theta^\text{cl}_k$, namely
\begin{equation}
    \dot{\theta}_k = -\frac{1}{S\sin\theta_k}\frac{\partial\vec{S}_k(t)}{\partial\phi_k}\cdot\sum_j\left[\left(\vec{\Gamma}_{kj}*\vec{S}_j\right)(t) + \vec{\xi}_j(t)\right] \label{eq:thetaeom1}
\end{equation}
and
\begin{equation}
    \dot{\phi}_k = -B_k + \frac{1}{S\sin\theta_k}\frac{\partial\vec{S}_k(t)}{\partial\theta_k}\cdot\sum_j\left[\left(\vec{\Gamma}_{kj}*\vec{S}_j\right)(t) + \vec{\xi}_j(t)\right] \label{eq:phieom1},
\end{equation}
where we now omitted the $\text{cl}$ superscript for the sake of conciseness. Following an approach similar to that which was used in Ref. \cite{Verstraten_PhysRevRes_5_033128_2023}, we show that these equations of motion are in fact exactly of the form of a generalised Landau-Lifshitz-Gilbert equation, namely
\begin{equation}
    \dot{\vec{S}}_k(t) = \vec{S}_k(t)\times\left[\vec{B}^\text{ext} + \sum_j(\vec{\Gamma}*\vec{S}_j)(t) + \vec{\xi}_j(t)\right]. \label{eq:llg1}
\end{equation}
It is well-known that the retarded and advanced components of the response function $\vec{\Pi}$ are related as \cite{Kamenev_FieldTheoryNonEq}
\begin{equation*}
    \left(\vec{\Pi}^R\right)^\dagger = \vec{\Pi}^A,
\end{equation*}
where $\dagger$ denotes complex conjugation and exchange of the time arguments, as well as the transpose with respect to both site and spin indices. One immediately concludes that the low-frequency expansion of $\vec{\Gamma}$ must give, in the frequency domain:
\begin{equation}
    \vec{\Gamma}(\omega) = \vec{K} + \ii\omega\vec{\alpha},
\end{equation}
where we defined the reactive and dissipative parts of the friction kernel as
\begin{align}
    \vec{K} &\equiv 2\lim_{\omega\to 0}\Re\left[\vec{\Pi}^R(\omega)\right]; \label{eq:fieldident}\\
    \vec{\alpha} &\equiv 2\lim_{\omega\to 0}\frac{\Im\left[\vec{\Pi}^R(\omega)\right]}{\omega}, \label{eq:dampingident}
\end{align}
such that Eq. (\ref{eq:llgnonsimp}) reduces to Eq. (\ref{eq:llg}), as claimed. \\
One now sees that, in order to make further progress in understanding the bath-induced dynamics of the spins, we need to proceed with evaluating the real and imaginary parts of $\vec{\Pi}^R$. This calculation forms the subject of the following section.
\section{Calculation of the response function} \label{sec:linres}
\subsection{Response functions in the frequency domain}
To make progress with the evaluation of the response function, we first extract its greater and lesser parts, and apply the following identities:
\begin{align}
    G^{\gtrless}(\vec{q, }\pm(t-t')) &= \frac{1}{2\pi}\int\rmd\omega \ \exp{\mp\ii\omega(t-t')}G^{\gtrless}(\vec{q},\omega) \label{eq:gfunctdom}; \\
    \ii G^>_s(\vec{q},\omega) &= A_s(\vec{q},\omega)\left(1 - N_\text{F}(\omega - \mu)\right); \label{eq:ggrfreqdom}\\
    \ii G^<_s(\vec{q},\omega) &= -A_s(\vec{q},\omega)N_\text{F}(\omega-\mu) \label{eq:glesfreqdom},
\end{align}
where $N_\text{F}(x) = 1 / (e^{\beta x} + 1)$ is the Fermi-Dirac distribution function, and $A_s(\vec{q},\omega)$ is the spectral function of electrons with momentum $\vec{q}$ and energy $\omega$ in the Rashba band $s$, defined by
\begin{equation}
    A_s(\vec{q},\omega) \equiv \ii\left[G^R_s(\vec{q},\omega) - G^A_s(\vec{q},\omega)\right]\label{eq:specfunc},
\end{equation}
which may be taken to have a $\delta$-function form if the finite lifetime of electrons in the bath is neglected:
\begin{equation}
     A_s(\vec{q},\omega) = 2\pi\delta(\epsilon_{\vec{q}s} -\omega). 
\end{equation}
Using the definitions from Eqs. (\ref{eq:radefs}-\ref{eq:kelddefs}), we obtain
\begin{widetext}
\begin{align}
    \left(\vec{\Pi}^{R(A)}\right)^{\mu\nu}_{ij}(\tau) &= \pm\Theta\left(\pm\tau\right)\frac{\ii J_{sd}^2}{2}\int\frac{\rmd\omega'}{2\pi}\int\frac{\rmd\omega''}{2\pi}\exp{-\ii(\omega' -\omega'')\tau}\left(N_\text{F}(\omega'-\mu) - N_\text{F}(\omega''-\mu)\right) \nonumber \\
    &\times\sum_{ss'}\int\frac{\rmd^2\vec{k}}{(2\pi)^2}\int\frac{\rmd^2\vec{k'}}{(2\pi)^2}\exp{-\ii(\vec{k} - \vec{k'})\cdot\vec{r}_{ij}}\mathcal{M}^{\mu\nu}_{ss'}(\vec{k},\vec{k'})A_{s}(\vec{k},\omega'')A_{s'}(\vec{k'},\omega') \\, 
    \left(\vec{\Pi}^K\right)^{\mu\nu}_{ij}(\tau) &= \frac{\ii J_{sd}^2}{2}\int\frac{\rmd\omega'}{2\pi}\int\frac{\rmd\omega''}{2\pi}\exp{-\ii(\omega' -\omega'')\tau}\left[N_\text{F}(\omega'' - \mu) + N_\text{F}(\omega' - \mu) - 2N_\text{F}(\omega'' - \mu)N_\text{F}(\omega' - \mu)\right] \nonumber \\
    &\times\sum_{ss'}\int\frac{\rmd^2\vec{k}}{(2\pi)^2}\int\frac{\rmd^2\vec{k'}}{(2\pi)^2}\exp{-\ii(\vec{k} - \vec{k'})\cdot\vec{r}_{ij}}\mathcal{M}^{\mu\nu}_{ss'}(\vec{k},\vec{k'})A_{s}(\vec{k},\omega'')A_{s'}(\vec{k'},\omega'), 
\end{align}
where we defined $\tau\equiv t-t'$ for brevity. The Fourier transform yields
\begin{align}
    \left(\vec{\Pi}^{R(A)}\right)^{\mu\nu}_{ij}(\omega) &= -\frac{J_{sd}^2}{2}\int\frac{\rmd\omega'}{2\pi}\int\frac{\rmd\omega''}{2\pi}\frac{1}{\omega - (\omega'-\omega'') \pm\ii 0}\left(N_\text{F}(\omega''-\mu) - N_\text{F}(\omega'-\mu)\right) \nonumber \\
    &\times\sum_{ss'}\int\frac{\rmd^2\vec{k}}{(2\pi)^2}\int\frac{\rmd^2\vec{k'}}{(2\pi)^2}\exp{-\ii(\vec{k} - \vec{k'})\cdot\vec{r}_{ij}}\mathcal{M}^{\mu\nu}_{ss'}(\vec{k},\vec{k'})A_{s}(\vec{k},\omega'')A_{s'}(\vec{k'},\omega'); \label{eq:retadfreqdom} \\
    \left(\vec{\Pi}^K\right)^{\mu\nu}_{ij}(\omega) &= \frac{\ii J_{sd}^2}{2}\int\frac{\rmd\omega'}{2\pi}\int\rmd\omega''\delta(\omega - (\omega' - \omega''))\left[N_\text{F}(\omega'' - \mu) + N_\text{F}(\omega' - \mu) - 2N_\text{F}(\omega'' - \mu)N_\text{F}(\omega' - \mu)\right] \nonumber \\
    &\times\sum_{ss'}\int\frac{\rmd^2\vec{k}}{(2\pi)^2}\int\frac{\rmd^2\vec{k'}}{(2\pi)^2}\exp{-\ii(\vec{k} - \vec{k'})\cdot\vec{r}_{ij}}\mathcal{M}^{\mu\nu}_{ss'}(\vec{k},\vec{k'})A_{s}(\vec{k},\omega'')A_{s'}(\vec{k'},\omega') \label{eq:keldfreqdom}
\end{align}
\end{widetext}
Before we proceed any further, we note an important relation between the Keldysh component and the retarded/advanced components. Specifically, one has:
\begin{equation}
    \left(\vec{\Pi}^K\right)^{\mu\nu}_{ij}(\omega) = \coth\left(\frac{\omega}{2T}\right)\left[\left(\vec{\Pi}^R\right)^{\mu\nu}_{ij}(\omega) - \left(\vec{\Pi}^A\right)^{\mu\nu}_{ij}(\omega)\right] \label{eq:fdt}.
\end{equation}
This equation holds generally whether one performs a low-frequency expansion of the response function or not, and constitutes a statement of the fluctuation-dissipation theorem: it relates the Keldysh component, determining the correlations of the \textit{fluctuating} magnetic fields $\vec{\xi}_i(t)$, to the imaginary part of $\vec{\Pi}^R$, which, as we shall see, determines the Gilbert damping, i.e. the dissipation. \\
We now continue with the calculation. First, we evaluate the momentum space integrals which appear in Eqs. (\ref{eq:retadfreqdom}--\ref{eq:keldfreqdom}). As shown in Appendix \ref{app:mspace}, these integrals have the following matrix structure:
\begin{widetext}
\begin{align}
    &\sum_{ss'}\int\frac{\rmd^2\vec{k}}{(2\pi)^2}\int\frac{\rmd^2\vec{k'}}{(2\pi)^2}\exp{-\ii(\vec{k} - \vec{k'})\cdot\vec{r}_{ij}}\mathcal{M}^{\mu\nu}_{ss'}(\vec{k},\vec{k'})A_{s}(\vec{k},\omega'')A_{s'}(\vec{k'},\omega') \\
    &= \frac{m^2}{2}\sum_{ss'}\frac{k_{0s}(\omega'')k_{0s'}(\omega')}{k_0(\omega')k_0(\omega'')}\mathds{J}^{\mu\nu}_{ss'}(\omega',\omega'';\vec{r}_{ij})\Theta(\omega'' - \epsilon_{0s})\Theta(\omega' - \epsilon_{0s'}) \label{eq:mspaceintfull},
\end{align}
 where the matrix $\mathds{J}^{\mu\nu}_{ss'}(\omega',\omega'';\vec{r}_{ij})$ has components
\begin{align*}
     \mathds{J}^{xx}_{ss'}(\omega',\omega'';\vec{r}_{ij}) &= J_0(k_{0s'}(\omega')R)J_0(k_{0s}(\omega'')R) - ss'\cos(2\theta)J_1(k_{0s'}(\omega')R)J_1(k_{0s}(\omega'')R); \\
     \mathds{J}^{yy}_{ss'}(\omega',\omega'';\vec{r}_{ij}) &=  J_0(k_{0s'}(\omega')R)J_0(k_{0s}(\omega'')R) + ss'\cos(2\theta)J_1(k_{0s'}(\omega')R)J_1(k_{0s}(\omega'')R); \\
     \mathds{J}^{zz}_{ss'}(\omega',\omega'';\vec{r}_{ij}) &=  J_0(k_{0s'}(\omega')R)J_0(k_{0s}(\omega'')R) - ss'J_1(k_{0s'}(\omega')R)J_1(k_{0s}(\omega'')R); \\
     \mathds{J}^{xy}_{ss'}(\omega',\omega'';\vec{r}_{ij}) &= \mathds{J}^{yx}_{ss'}(\omega',\omega'';\vec{r}_{ij}) = -\frac{ss'}{2}\sin(2\theta) J_1(k_{0s'}(\omega')R)J_1(k_{0s}(\omega'')R); \\
     \mathds{J}^{xz}_{ss'}(\omega',\omega'';\vec{r}_{ij}) &= -\mathds{J}^{zx}_{ss'}(\omega',\omega'';\vec{r}_{ij}) = -\left(s'J_1(k_{0s'}(\omega')R)J_0(k_{0s}(\omega'')R)+sJ_0(k_{0s'}(\omega')R)J_1(k_{0s}(\omega'')R)\right)\cos\theta ;\\
     \mathds{J}^{yz}_{ss'}(\omega',\omega'';\vec{r}_{ij}) &= -\mathds{J}^{zy}_{ss'}(\omega',\omega'';\vec{r}_{ij}) = \left(s'J_1(k_{0s'}(\omega')R)J_0(k_{0s}(\omega'')R)+sJ_0(k_{0s'}(\omega')R)J_1(k_{0s}(\omega'')R)\right)\sin\theta,
 \end{align*}
\end{widetext}
where $\vec{r}_{ij} = (R\cos\theta, R\sin\theta)$, $J_n(x)$ is the Bessel function of the first kind at order $n$ and we defined
\begin{align*}
    k_0(\epsilon) &= \sqrt{q_R^2 + 2m\epsilon}, \\
    k_{0s}(\epsilon) &= -sq_R + k_0(\epsilon).
\end{align*}
The step functions appearing in Eq. (\ref{eq:mspaceintfull}) are necessary to enforce the support of the spectral function $A_{\alpha}(\vec{k},\omega)$ on the physical domain $k > 0$ for all values of $\omega$. \\
\subsection{Keldysh component}
Our aim is to perform a low-frequency expansion of the expressions (\ref{eq:retadfreqdom}, \ref{eq:keldfreqdom}). For the retarded and advanced components, we expand to first order in $\omega$, whereas for the Keldysh component it is sufficient to expand to zeroth order.\\
We first evaluate the latter, which contains a $\delta$ function that sets $\omega'' \to \omega' - \omega$. Before proceeding further, we examine the effect that this has on the integrand. Focusing on the term containing Fermi functions, one can show that, after integrating over $\omega''$,
\begin{align}
    &\left[N_\text{F}(\omega'' - \mu) + N_\text{F}(\omega' - \mu) - 2N_\text{F}(\omega'' - \mu)N_\text{F}(\omega' - \mu)\right]_{\omega=0} \nonumber \\
    &\approx 2  T\delta(\omega'-\mu),
\end{align}
in the limit that the derivative of the Fermi function with respect to the Fermi level $\mu$ can be approximated as a $\delta$-function; this approximation is excellent provided $T \ll T_F$, and is fulfilled experimentally in a variety of archetypal 2DEGs \cite{LaShell1996_Au111_Rashba_PRL,Reinert2001_Lgap_CuAgAu_PRB,Nicolay2001_AuAg111_SpinOrbit_PRB,Hoesch2004_Au111_SpinStructure_PRB,Tamai2013_Cu111_Rashba_PRB,Hufner2008_Lgap_NobleMetals_ZPhyChem,Ast2007_BiAg111_GiantRashba_PRL,Gierz2010_StructuralInfluence_Rashba_ArXiv,Meier2009_BiPbAg_TunableFermi_PRB,Pacile2006_PbAg111_PRB,Varykhalov2012_Ir111_GiantRashba_PRL}. \\
\begin{widetext}
The Keldysh component becomes, to zeroth order in $\omega$,
\begin{align} \label{eq:keldlowfreq}
    \left(\vec{\Pi}^K\right)^{\mu\nu}_{ij}(\omega) &\approx \frac{\ii T J_{sd}^2m^2}{4\pi}\sum_{ss'}\left(1-s\frac{sq_R}{k_F}\right)\left(1-\frac{s'q_R}{k_F}\right)\mathds{J}^{\mu\nu}_{ss'}(\mu,\mu;\vec{r}_{ij}), 
\end{align}
where we defined the Fermi wavevector $k_F = \sqrt{2m\mu + q_R^2} = k_0(\mu)$ and $k_{Fs} = -sq_R + k_F$. We now proceed with the calculation of the retarded and advanced components.
\subsection{Retarded and advanced components}
We first split the energy denominator in Eq. (\ref{eq:retadfreqdom}) into its real and imaginary parts according to the Sokhotski–Plemelj theorem \cite{Weinberg_QFT1, Plemelj_RiemannKlein_1964}, and substitute Eq. (\ref{eq:mspaceintfull}) for the momentum space integrals, yielding:
\begin{align}
    &\left(\vec{\Pi}^{R(A)}\right)^{\mu\nu}_{ij}(\omega) = -\frac{J_{sd}^2m^2}{4}\sum_{ss'}\mathcal{P}\int_{\epsilon_{0s'}}^\infty\frac{\rmd\omega'}{2\pi}\int_{\epsilon_{0s}}^\infty\frac{\rmd\omega''}{2\pi}\frac{\left(N_\text{F}(\omega''-\mu) - N_\text{F}(\omega'-\mu)\right)}{\omega - (\omega'-\omega'')}\frac{k_{0s}(\omega'')k_{0s'}(\omega')}{k_0(\omega')k_0(\omega'')}\mathds{J}^{\mu\nu}_{ss'}(\omega',\omega'';\vec{r}_{ij}) \nonumber \\
    \label{eq:pirareal} \\
    &\pm\frac{\ii\pi J_{sd}^2m^2}{4}\sum_{ss'}\int_{\epsilon_{0s'}}^\infty\frac{\rmd\omega'}{2\pi}\int_{\epsilon_{0s}}^\infty\frac{\rmd\omega''}{2\pi}\delta(\omega - (\omega'-\omega''))\left(N_\text{F}(\omega''-\mu) - N_\text{F}(\omega'-\mu)\right)\frac{k_{0s}(\omega'')k_{0s'}(\omega')}{k_0(\omega')k_0(\omega'')}\mathds{J}^{\mu\nu}_{ss'}(\omega',\omega'';\vec{r}_{ij})  \label{eq:piraimag},
\end{align}
where $\mathcal{P}$ means that the result of the integral is to be understood in the Cauchy principal value sense, and we set the lower bounds of the integrals using the step functions in Eq. (\ref{eq:mspaceintfull}). \\ 
The frequency integrals in Eq. (\ref{eq:piraimag}) are easily evaluated: we first use the $\delta$-function to evaluate the integral over $\omega''$, setting $\omega''\to\omega'-\omega$ as before. This shows that the zeroth order term in the low frequency expansion of (\ref{eq:piraimag}) vanishes when evaluated at $\omega=0$. The first order term is easily evaluated using integration by parts. We find that
\begin{align}
    \Im\left[\left(\vec{\Pi}^{R(A)}\right)^{\mu\nu}_{ij}(\omega)\right] &\approx \frac{\pm\omega J_{sd}^2m^2}{16\pi}\sum_{ss'}\frac{k_{Fs}k_{Fs'}}{k_F^2}\mathds{J}^{\mu\nu}_{ss'}(\mu,\mu;\vec{r}_{ij}) . \label{eq:raimaglowfreq}
\end{align}
It is this imaginary component of the retarded response function which enters into the Gilbert damping in the equation of motion (\ref{eq:llg}). Applying the identity (\ref{eq:dampingident}), we find for the Gilbert damping tensor:
\begin{align}
    \vec{\alpha}^{\mu\nu}_{ij} &= 2\lim_{\omega\to 0}\frac{1}{\omega}\Im\left[\left(\vec{\Pi}^{R}\right)^{\mu\nu}_{ij}(\omega)\right] \nonumber \\
    &= \frac{J_{sd}^2m^2}{8\pi}\sum_{ss'}\frac{k_{Fs}k_{Fs'}}{k_F^2}\mathds{J}^{\mu\nu}_{ss'}(\mu,\mu;\vec{r}_{ij}) \nonumber \\
    &= \frac{\alpha}{4} \sum_{ss'}\left(1-\frac{sq_R}{k_F}\right)\left(1-\frac{s'q_R}{k_F}\right)\mathds{J}^{\mu\nu}_{ss'}(\mu,\mu;\vec{r}_{ij}) \label{eq:alphatensor},
\end{align}
where the local damping $\alpha$ is the limit of Eq. (\ref{eq:alphatensor}) as $R \to 0$ and is given by
\begin{equation}
    \alpha = \frac{J_{sd}^2m^2}{2\pi}. \label{eq:alphaloc}
\end{equation}
The structure of Eq. (\ref{eq:alphatensor}) reproduces the expression which was recently derived for the Gilbert damping at a heavy metal/ferromagnet interface \cite{OsorioNonLocal}. \\
Before proceeding with the principal value parts, we observe that
\begin{equation*}
    \left(\vec{\Pi}^K\right)^{\mu\nu}_{ij}(\omega) = \frac{2T}{\omega}\left[\left(\vec{\Pi}^R\right)^{\mu\nu}_{ij}(\omega) - \left(\vec{\Pi}^A\right)^{\mu\nu}_{ij}(\omega)\right],
\end{equation*}
which is the low-frequency version of the FDT given in Eq. (\ref{eq:fdt}). Using this result, and recalling Eqs. (\ref{eq:xicorrels}) and (\ref{eq:dampingident}), we find that, in the time domain,
\begin{equation}
    \Braket{\vec{\xi}_i(t)\vec{\xi}_j(t')} = T\vec{\alpha}_{ij}\delta(t-t').
\end{equation}
For the principal value part (\ref{eq:pirareal}), we may not trivially expand the denominator in powers of $\omega$. The reason for this is that, when we take the static limit $\omega \to 0$ as required by Eq. (\ref{eq:fieldident}), the integral will diverge on the boundary $\omega' = \omega'' = \mu$, making the principal value ill-defined in this case. To proceed, we must make two important assumptions. We firstly assume that the Bessel functions appearing in Eq. (\ref{eq:mspaceintfull}) can be replaced by their asymptotic forms. Secondly, observing the structure of the denominator (and keeping in mind that we later take the limit $\omega \to 0$), we can expand the functions $k_0(\omega'')$ and $k_0(\omega')$ around $\omega' = \omega'' = \mu$, since this is the largest contribution to the integrand. More details of the validity of these approximations and the full calculation are given in Appendix \ref{app:mspace}. For now, we give results for the case where $\theta = 0$ (i.e. $\vec{r}_{12} = \hat{\vec{x}}$)
\begin{align}
    \lim_{\omega\to 0}\Re\left[\left(\vec{\Pi}^R(\omega)\right)^{\mu\nu}_{ij}\right] = \frac{J_{sd}^2mk_F^2}{8\pi^2}F(k_FR)\mathds{M}(q_RR; \theta) \label{eq:realpartstatic},
\end{align}
where the range function $F(k_FR)$ and matrix $\mathds{M}(q_RR; \theta)$ are given by 
\begin{equation}
    F(k_FR) = \frac{\sin(2k_FR)}{(k_FR)^2}, \label{eq:rangefunc}
\end{equation}
and
\begin{equation}
    \mathds{M}(q_RR; \theta) = \begin{pmatrix}
        1 - \cos(2\theta) + \cos(2q_RR)(1 + \cos(2\theta)) & \frac{1}{2}\sin(2\theta)(\cos(2q_RR) - 1) & \cos\theta\sin(2q_RR) \\
        \frac{1}{2}\sin(2\theta)(\cos(2q_RR) - 1) & 1 + \cos(2\theta) + \cos(2q_RR)(1 - \cos(2\theta)) & -\sin\theta\sin(2q_RR) \\
        -\cos\theta\sin(2q_RR) & \sin\theta\sin(2q_RR) & 2\cos(2q_RR)
    \end{pmatrix},
\end{equation}
respectively. The exchange tensor is then trivially deduced from Eq. (\ref{eq:fieldident}). This is
\begin{equation}
    \vec{K}^{\mu\nu}_{ij} = \frac{J_{sd}^2mk_F^2}{4\pi^2}F(k_FR)\mathds{M}(q_RR; \theta) \label{eq:Ktensor}.
\end{equation}
For brevity, we here define the constant prefactor appearing in Eq. (\ref{eq:Ktensor}) as
\begin{equation}
    K = \frac{J_{sd}^2mk_F^2}{4\pi^2}. \label{eq:Kprefactor}
\end{equation}
\\
Having derived expressions for the various components of the response function, we proceed to further simplification of the equations of motion (\ref{eq:llg}).
\end{widetext}
\section{Generalised Landau-Lifshitz-Gilbert equation} \label{sec:llg}
\begin{figure}
    \begin{center}
    \includegraphics[width=\linewidth]{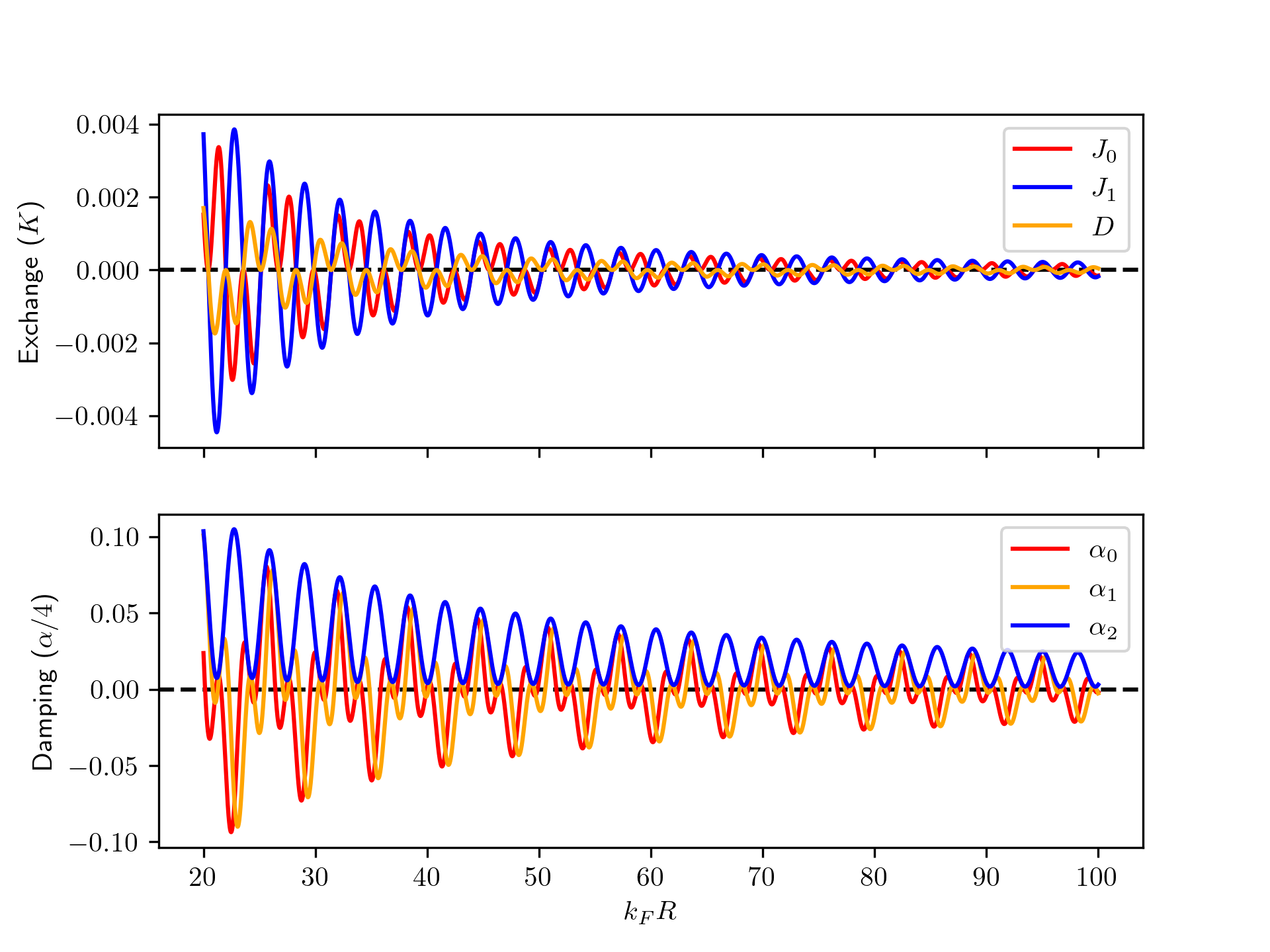}
    \end{center}
    \caption{Exchange and damping parameters in Eqs. (\ref{eq:exchnspec}--\ref{eq:dampspec}), normalised by the respective prefactors in Eqs. (\ref{eq:alphatensor}) and (\ref{eq:Ktensor}).}
    \label{fig:exanddamp}
\end{figure}

We begin by expanding the equation of motion (\ref{eq:llg}) for each spin:
\begin{align}
    \dot{\vec{S}}_1 &= \vec{S}_1\times\vec{B}^\text{ext} - \alpha\vec{S}_1\times\dot{\vec{S}}_1 \nonumber \\
    &+ \vec{S}_1\times\left[\vec{K}_{12}\vec{S}_2 - \vec{\alpha}_{12}\dot{\vec{S}}_2\right] \label{eq:llgs1} \\
    \dot{\vec{S}}_2 &= \vec{S}_2\times\vec{B}^\text{ext} - \alpha\vec{S}_2\times\dot{\vec{S}}_2 \nonumber \\
    &+ \vec{S}_2\times\left[\vec{K}_{21}\vec{S}_1 - \vec{\alpha}_{21}\dot{\vec{S}}_1\right], \label{eq:llgs2}
\end{align}
dropping stochastic field terms because these are not important for the non-reciprocal physics that we are interested in. \\
One observes that the equations of motion contain no local exchange terms of the form $\vec{K}_{11}$ or $\vec{K}_{22}$. These are absent because, as shown in Appendix \ref{app:mspace}, they are proportional to the identity, and hence contribute no torque to the LLG. Physically, this is expected; a Rashba 2DEG has an isotropic Fermi surface, prohibiting local anisotropy in a spin polarisation which is induced by a magnetic impurity. \\
Now, choosing a particular geometry such that $\vec{r}_{12} = R\hat{\vec{x}} = -\vec{r}_{21}$, we find that the DMI is parallel to $\hat{\vec{y}}$. In this geometry, the exchange tensors $\vec{K}_{12} = \vec{K}_{21}^T$ and $\vec{\alpha}_{12} = \vec{\alpha}_{21}^T$ have the following structure:
\begin{align}
    \vec{K}_{12} &= \vec{K}^T_{21} = \begin{pmatrix}
        J_0 & 0 & D \\ 
        0 & J_1 & 0 \\
        -D & 0 & J_0
    \end{pmatrix}; \label{eq:exchnspec}
\end{align}
\begin{align}
        \vec{\alpha}_{12} &= \vec{\alpha}^T_{21} = \begin{pmatrix}
        \alpha_0 & 0 & -\alpha_1 \\ 
        0 & \alpha_2 & 0 \\
        \alpha_1 & 0 & \alpha_0
    \end{pmatrix} \label{eq:dampspec}.
\end{align}
Before proceeding, we briefly discuss the structure of Eqs. (\ref{eq:exchnspec}--\ref{eq:dampspec}). One observes that these (i): are anisotropic in space and (ii): contain antisymmetric contributions. In the case of the exchange tensor, this antisymmetric contribution is the DMI, whereas in the case of the damping tensor it corresponds to chiral damping. The spin-orbit coupling of the bath, in combination with the broken inversion symmetry which is induced by the presence of the spins on the surface of the former, leads to this anisotropic, antisymmetric structure \cite{PhysRev.120.91,DZYALOSHINSKY1958241}. \\
The quantities $J_0$, $J_1$, $D$ and $\alpha_0$, $\alpha_1$, $\alpha_2$ may be deduced from Eqs. (\ref{eq:alphatensor}) and (\ref{eq:realpartstatic}), respectively. These quantities are plotted in Fig. \ref{fig:exanddamp} for the ratio $q_R / k_F = 0.5$. In the following sub--section, we linearise the equations of motion (\ref{eq:llgs1}-\ref{eq:llgs2}) and examine the magnon susceptibility.
\subsection{Linearised dynamics and non-reciprocity}
\begin{figure}
    \begin{center}
    \includegraphics[width=\linewidth]{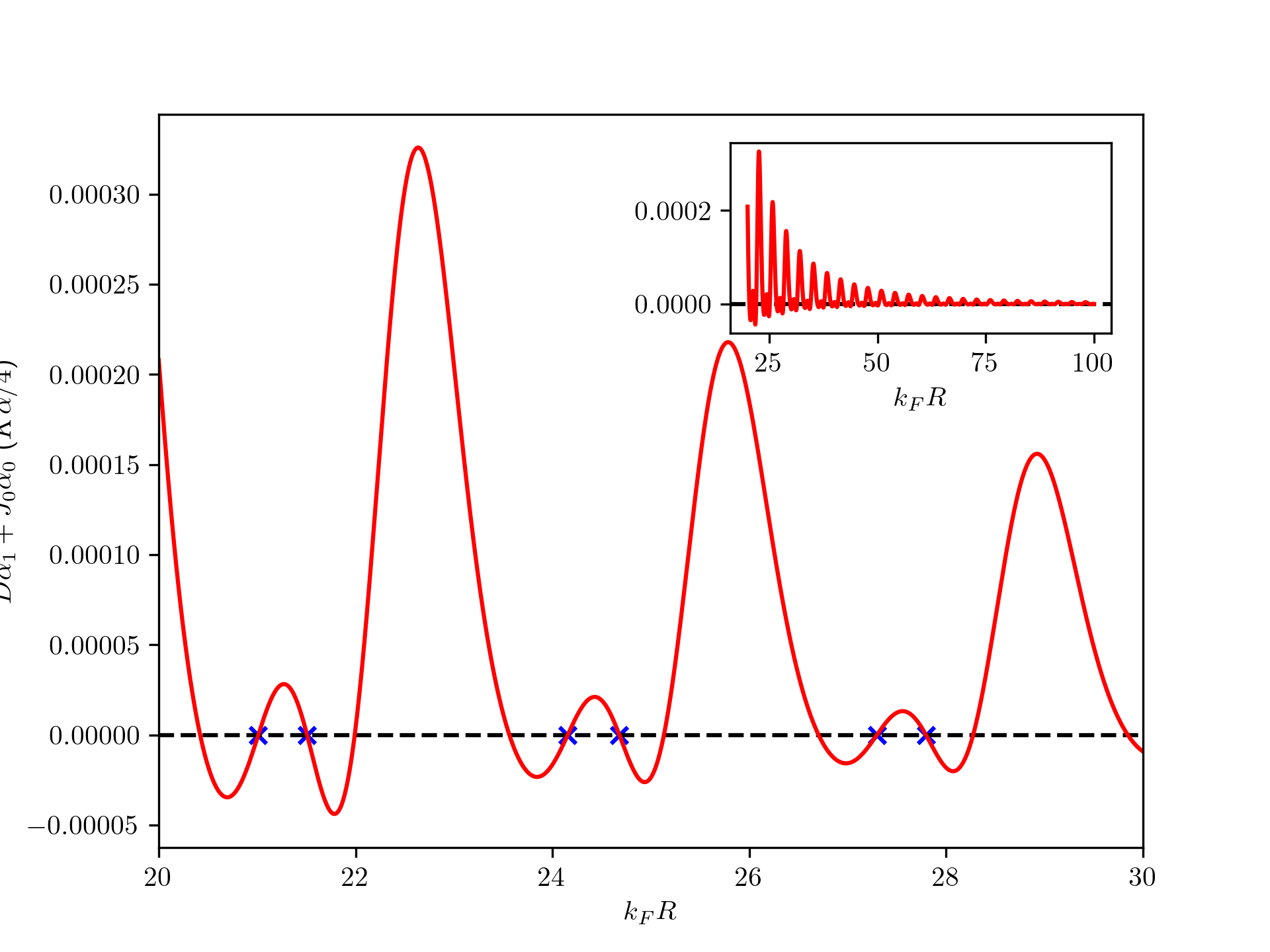}
    \end{center}
    \caption{Condition in Eq. (\ref{eq:finetuning}) as a function of $k_FR$, with non-trivial roots plotted as blue crosses. The inset plot depicts the variation of this quantity across the same range of $k_FR$ as depicted in Fig. \ref{fig:exanddamp}.}
    \label{fig:finetuning}
\end{figure}

\begin{figure}
    \begin{center}
    \includegraphics[width=\linewidth]{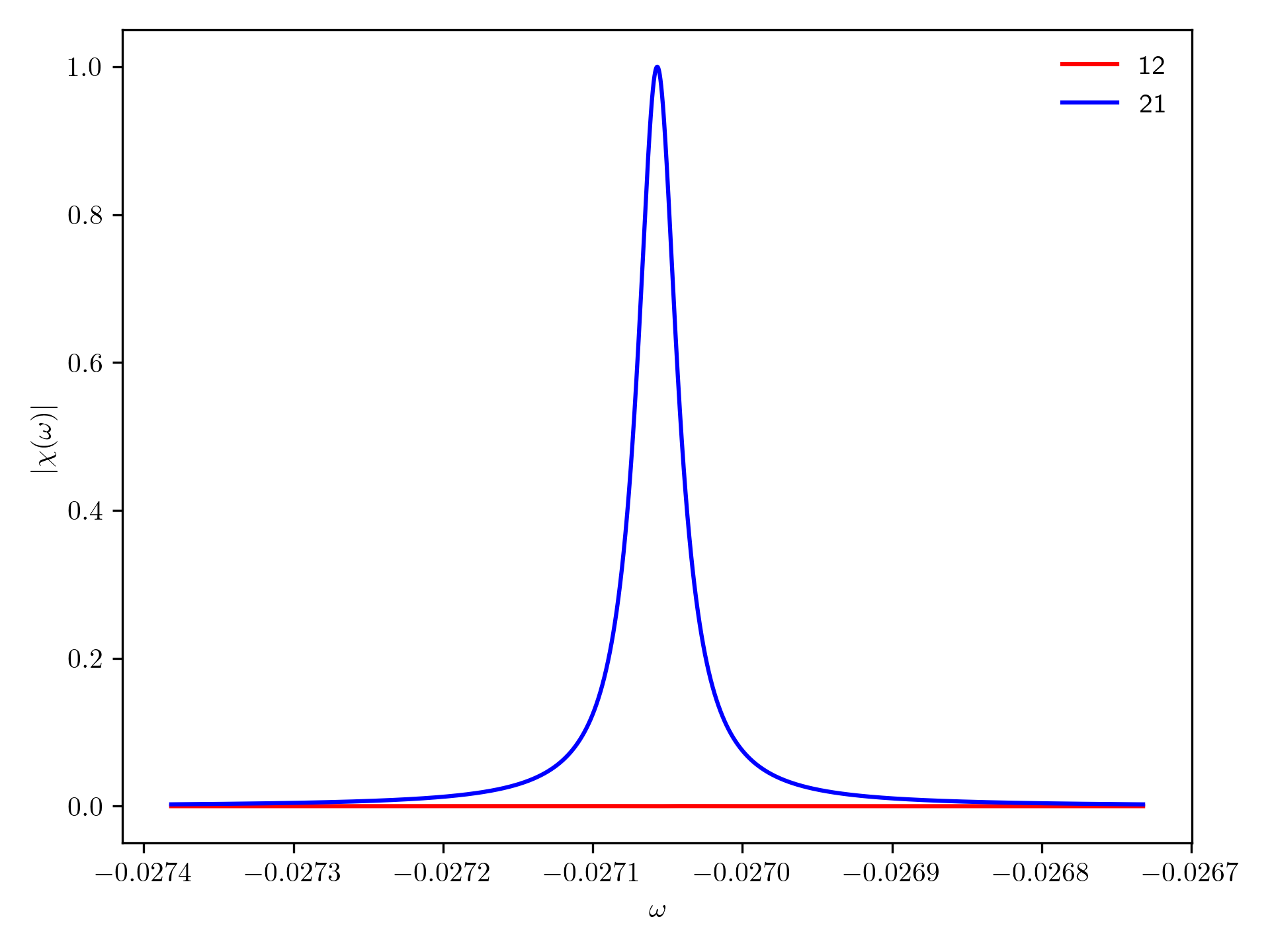}
    \end{center}
    \caption{Normalised magnon susceptibilities $\chi_{12}$ and $\chi_{21}$ with the condition (\ref{eq:x12cancels}).}
    \label{fig:freqnonrecip}
\end{figure}

We now proceed to study the linearisation of the equations of motion (\ref{eq:llgs1}-\ref{eq:llgs2}). Since the DMI is parallel to $\hat{\vec{y}}$, we follow Ref. \cite{Yuan_PhysRevB_107_024418_2023}, taking $\vec{B}^\text{ext} = B_0\hat{\vec{y}}$, and linearise Eqs. (\ref{eq:llgs1}--\ref{eq:llgs2}) around this direction. For the sake of convention, we first rotate (\ref{eq:llgs1}--\ref{eq:llgs2}) into a frame such that $\hat{\vec{y}} \to\hat{\vec{z}}$. Then, in this rotated frame, we consider small transverse deviations $\delta S_{ix}, \delta S_{iy}$ in the $(\hat{\vec{x}}, \hat{\vec{y}})$ plane. Projecting the equations of motion onto this plane, transforming to magnon coordinates $\psi_i = \delta S_{ix} + \ii\delta S_{iy}$ and taking the Fourier transform, we find that Eqs. (\ref{eq:llgs1}--\ref{eq:llgs2}) reduce to
\begin{equation}
    \chi^{-1}(\omega)\begin{pmatrix}
        \psi_1 \\
        \psi_2
    \end{pmatrix} = 0,
\end{equation}
where the magnetic susceptibility, or magnon Green's function $\chi$ is given by
\begin{equation}
    \chi(\omega) = \frac{1}{\det(\chi^{-1}(\omega))}\begin{pmatrix}
        \chi_{11}(\omega) & \chi_{12}(\omega) \\
        \chi_{21}(\omega) & \chi_{11}(\omega)
    \end{pmatrix}.
\end{equation}
Here,
\begin{align*}
    \chi_{11}(\omega) &= \ii\omega(1 - \ii\alpha S) + \ii(B_0 - SJ_1); \\
    \chi_{12}(\omega) &= -S(\ii\omega(\alpha_1 - \ii\alpha_0) + (D - \ii J_0));\\
    \chi_{21}(\omega) &= S(\ii\omega(\alpha_1 + \ii\alpha_0) + (D + \ii J_0));\\
    \det(\chi^{-1}(\omega)) &= \chi_{11}^2(\omega) - \chi_{12}(\omega)\chi_{21}(\omega).
\end{align*}
Taking a weak-damping limit such that $\alpha, \alpha_0, \alpha_1 \ll 1$, one finds that the resonance frequencies, determined by solving $\det(\chi^{-1}(\omega)) = 0$, are
\begin{equation}
    \omega_{r\pm} = \mp S\Omega_0 -\omega_0  \ + \frac{\ii S}{\Omega_0}(S\Omega_0 \pm \omega_0)\left(D\alpha_1 + J_0\alpha_0 \mp \alpha\Omega_0\right),
\end{equation}
where $\omega_0 = B_0 - SJ_1$ and $\Omega_0 = \sqrt{D^2 + J_0^2}$. \\
Evaluating the off-diagonal elements $\chi_{12}$ and $\chi_{21}$ at these resonance frequencies, and again dropping terms of second order or higher in the dampings, we find:
\begin{align*}
    \chi_{12}(\omega_{r\pm}) &= S(-D + \alpha_0(\omega_0 \pm S\Omega_0)) \nonumber \\
    &+ \ii S(J_0 + \alpha_1(\omega_0 \pm S\Omega_0)); \\
    \chi_{21}(\omega_{r\pm}) &= S(D + \alpha_0(\omega_0 \pm S\Omega_0)) \\
    &+ \ii S(J_0 -\alpha_1(\omega_0 \pm S\Omega_0)).
\end{align*}
Now, true unidirectional transmission of magnons from spin $1$ to spin $2$ (for example) corresponds to $\chi_{12}(\omega_{r\pm}) = 0$ while $\chi_{21}(\omega_{r\pm}) \neq 0$. The parameter which can be controlled to satisfy this condition is the strength of the external field $B_0$, which enters into the term $\omega_0$. Hence, we solve for a condition on $\omega_0$ such that the real and imaginary parts are simultaneously zero. The resulting condition on $B_0$ is given by
\begin{align}
    \chi_{12}(\omega_{r\pm}) = 0 \ \text{at} \ B_{0\pm} &= \frac{D}{\alpha_0} + S(J_1 \mp\Omega_0)   \label{eq:x12cancels} \\
    \chi_{21}(\omega_{r\pm}) = 0 \ \text{at} \ B_{0\pm} &= -\frac{D}{\alpha_0} + S(J_1 \mp\Omega_0),\label{eq:x21cancels}
\end{align}
provided we also impose an additional condition on the parameters
\begin{equation}
    D\alpha_1 + J_0\alpha_0 = 0 \label{eq:finetuning},
\end{equation}
for both cases. In a generic system, there are always certain interatomic spacings $R$ at which the condition(s) are satisfied. For the same ratio $q_R / k_F = 0.5$ as was used in Fig. \ref{fig:exanddamp}, we plot Eq. (\ref{eq:finetuning}) in Fig. \ref{fig:finetuning}; one observes that there are indeed numerous crossings of the horizontal axis, confirming that the condition (\ref{eq:finetuning}) can be satisfied. \\
Taking the first non-trivial root of this quantity in the range depicted in Figs. \ref{fig:exanddamp} and \ref{fig:finetuning}, which occurs at $k_FR \approx 21.114$, and reasonable values for the parameters $K$ and $\alpha$ of $K \sim 10\mu$eV, $\alpha \sim 10^{-4}$, we compute the relevant magnetic parameters in Eqs. (\ref{eq:exchnspec}--\ref{eq:dampspec}) and plot the resulting magnon transmission for the field $B_{0+}$ in Fig. \ref{fig:freqnonrecip}. One observes from this figure that the magnon transmission from spin $2$ to spin $1$ has a peak at $\omega_{r+}$, whereas the transmission from spin $1$ to spin $2$ is exactly zero at the same frequency, confirming that the transmission is unidirectional.
\newpage
\section*{Conclusion}
In this work, we developed a model for an atomic spin diode, consisting of two localised spins (magnetic adatoms), deposited on the surface of a two-dimensional electron gas. We derived the semi--classical equations of motion for the steady-state spin dynamics using the Keldysh path integral formalism, and showed that these are very similar in form to the phenomenological equations which were the starting point of Ref. \cite{Yuan_PhysRevB_107_024418_2023}. Our microscopic approach gives exact expressions for both the local and non--local magnetic exchange and Gilbert damping parameters which enter into these equations of motion. Under certain assumptions, these all have analytical expressions. \\
The results we obtained rest on several key assumptions and approximations. Firstly, we treated the \textit{s-d} coupling constant perturbatively in order to integrate out the bath. Further, we assumed that the temperature of the system is much smaller than the Fermi temperature of the bath. These first two are the only two approximations that are necessary to arrive at an analytical expression for the damping; their validity is discussed in Sec. \ref{sec:effaction} and \ref{sec:linres}, respectively. \\
To obtain the quoted expression for the exchange tensor, we needed to make two further approximations in order to evaluate the frequency integrals in Eq. (\ref{eq:pirareal}). The validity of these approximations is discussed in Appendix \ref{app:mspace}. \\
As was emphasised in Ref. \cite{Yuan_PhysRevB_107_024418_2023}, a system which possesses both DMI, long--known to give non-reciprocal effects in spin-wave physics, and a non-local Gilbert damping, can exhibit unidirectional transmission of spin between the two magnetic subsystems, provided symmetric exchange is absent. To this end, we also studied the magnetic susceptibility, which is obtained by linearising the equations of motion and taking the Fourier transform. By tuning of the exchange and damping parameters, and carefully choosing the strength of the applied magnetic field, we find that the transmission of angular momentum between spin $1$ and spin $2$ (or vice versa) can always be made unidirectional. \\
In constrast to the work of Ref. \cite{Yuan_PhysRevB_107_024418_2023}, the symmetric exchange (denoted as $J_0$ in this work) is generally non-zero, and we also have a chiral damping $\alpha_1$, which was absent in the work of Yuan. \textit{et. al.}. Nevertheless, we were able to derive two conditions on both the external magnetic field $B_0$ and the interatomic spacing $R$ to realise such unidirectionality. The oscillatory natures of the magnetic parameters we calculated provide for high tunability, allowing our derived conditions to be satisfied independent of the underlying microscopic properties of the bath, such as the Fermi energy, Rashba spin splitting, and effective electron mass; these quantities only determine the relative magnitudes of the effective magnetic parameters and do not have an impact on the existence of the conditions (\ref{eq:x12cancels}--\ref{eq:finetuning}). \\ 
\\
Our work provides the first derivation, starting from a microscopic model, of the conditions for unidirectional spin-wave propagation which were laid down phenomenologically in Ref. \cite{Yuan_PhysRevB_107_024418_2023}. We emphasise that the conditions we derived are generic, though for the experimental parameters that we were able to infer, we found that the magnetic fields required to satisfy the conditions (\ref{eq:x12cancels}--\ref{eq:x21cancels}) are currently too large to realise experimentally \cite{Madhavan_PhysRevB_64_165412_2001,Reinert2001_Lgap_CuAgAu_PRB,Nicolay2001_AuAg111_SpinOrbit_PRB,Hoesch2004_Au111_SpinStructure_PRB,Madhavan1998ScienceKondo,Knorr2002PRL096804,Jamneala2000PRB9990,Limot2005PRL036805}. Nevertheless, the field of magnetic adatoms on surfaces is a very active one, and we hope that our work motivates further study of experimental setups in which realising these conditions is more feasible. We also foresee several other extensions to our work, which could include: going beyond the semi--classical approximation, using a more realistic form of the s-d interaction, explicitly considering the crystalline structure of the electronic substrate, or considering a system of more than two spins. \\
\section*{Acknowledgements}
W.J.H acknowledges the research program “Materials for the Quantum Age” (QuMat) for financial support. This program (registration number 024.005.006) is part of the Gravitation program financed by the Dutch Ministry of Education, Culture and Science (OCW).
\appendix
\begin{widetext}
\section{Derivation of Eqs. (\ref{eq:alphatensor}) and (\ref{eq:Ktensor})} \label{app:mspace}
We begin by re-quoting Eq. (\ref{eq:retadfreqdom}) from the main text:
\begin{align*}
    \left(\vec{\Pi}^{R(A)}\right)^{\mu\nu}_{ij}(\omega) &= -\frac{J_{sd}^2}{2}\int\frac{\rmd\omega'}{2\pi}\int\frac{\rmd\omega''}{2\pi}\frac{1}{\omega - (\omega'-\omega'') \pm\ii 0}\left(N_\text{F}(\omega''-\mu) - N_\text{F}(\omega'-\mu)\right) \nonumber \\
    &\times\sum_{ss'}\int\frac{\rmd^2\vec{k}}{(2\pi)^2}\int\frac{\rmd^2\vec{k'}}{(2\pi)^2}\exp{-\ii(\vec{k} - \vec{k'})\cdot\vec{r}_{ij}}\mathcal{M}^{\mu\nu}_{ss'}(\vec{k},\vec{k'})A_{s}(\vec{k},\omega'')A_{s'}(\vec{k'},\omega');  \\
    \left(\vec{\Pi}^K\right)^{\mu\nu}_{ij}(\omega) &= \frac{\ii J_{sd}^2}{2}\int\frac{\rmd\omega'}{2\pi}\int\rmd\omega''\delta(\omega - (\omega' - \omega''))\left[N_\text{F}(\omega'' - \mu) + N_\text{F}(\omega' - \mu) - 2N_\text{F}(\omega'' - \mu)N_\text{F}(\omega' - \mu)\right] \nonumber \\
    &\times\sum_{ss'}\int\frac{\rmd^2\vec{k}}{(2\pi)^2}\int\frac{\rmd^2\vec{k'}}{(2\pi)^2}\exp{-\ii(\vec{k} - \vec{k'})\cdot\vec{r}_{ij}}\mathcal{M}^{\mu\nu}_{ss'}(\vec{k},\vec{k'})A_{s}(\vec{k},\omega'')A_{s'}(\vec{k'},\omega') .
\end{align*}
\subsection{Evaluating the momentum space integrals}
The momentum space integrals are of the form
\begin{equation} \label{eq:mspaceint}
    \sum_{ss'}\int\frac{\rmd^2\vec{k}}{(2\pi)^2}\int\frac{\rmd^2\vec{k'}}{(2\pi)^2}\exp{-\ii(\vec{k} - \vec{k'})\cdot\vec{r}_{ij}}\mathcal{M}^{\mu\nu}_{ss'}(\vec{k},\vec{k'})A_{s}(\vec{k},\omega'')A_{s'}(\vec{k'},\omega').
\end{equation}
To proceed, we first rewrite the spectral function somewhat. As defined in Eq. (\ref{eq:specfunc}) of the main text, this is given by
\begin{equation*}
    A_{s}(\vec{k},\omega) = 2\pi\delta\left(\epsilon_{\vec{k}s}- \omega\right).
\end{equation*}
We wish to employ the distributional identity
\begin{equation*}
    \delta(f(k)) = \sum_{k_{0i}}\frac{\delta(k - k_{0i})}{f'(k_{0i})}, 
\end{equation*}
where $\left\{k_{0i}\right\}$ are the roots of $f(k) = 0$, to rewrite the spectral function in terms of $k$. Identifying $f(k) = \epsilon_{\vec{k}s} - \hbar\omega$, and taking note of the domain $k\geq 0$, we find one real root:
\begin{equation*}
    k_{0\alpha}(\omega) = -s q_R + \sqrt{q_R^2 + 2m\omega},
\end{equation*}
whereas the derivative of $f(k)$ is
\begin{equation*}
    f'(k) = \frac{1}{m}(k + \alpha q_R),
\end{equation*}
which, evaluated at the roots $k_{0\alpha}(\omega)$, gives
\begin{equation*}
    f'(k_{0\alpha}(\omega)) = \frac{1}{m}\sqrt{q_R^2 + 2m\omega} \equiv \frac{k_0(\omega)}{m}
\end{equation*}
where $q_R = m\soconst $, and we discarded the negative branch of the square root because the physical domain of $k$ is $k \geq  0$. It is also important to consider the support of the spectral function. In order for this to be well-defined in the distributional sense, we must choose $\omega$ such that the function $f(k)$ has real roots on the whole domain of $k$. Differentiation of $\epsilon_{\vec{k}\alpha}$ yields
\begin{equation*}
    \frac{\partial\epsilon_{\vec{k}s}}{\partial k} = \frac{k}{m} + s\lambda_{so} = 0.
\end{equation*}
For the $s = +1$ branch, we find no solutions for $k > 0$. For the $s = -1$ branch, we find a minimum at $k = m\soconst = q_R$. Hence
\begin{equation*}
    \epsilon_{0s} = \begin{cases}
        0 & \ s = +1 \\
        -\frac{q_R^2}{2m} & \ s= -1.
    \end{cases}
\end{equation*}
Therefore, in rewriting the spectral function, we must include a factor of $\Theta(\omega - \epsilon_{0s})$. To summarise, we have
\begin{equation*}
    A_{s}(\vec{k},\omega) = 2\pi m\frac{1}{k_0(\omega)}\delta(k - k_{0s}(\omega))\Theta(\omega - \epsilon_{0s}) .
\end{equation*}
We know that
\begin{equation*}
    \mathcal{M}^{\mu\nu}_{\vec{k}\vec{k'},ss'} = \Braket{s,\vec{k}|\sigma^\mu|s',\vec{k'}}\Braket{s',\vec{k'}|\sigma^\nu|s,\vec{k}}
\end{equation*}
where the helicity eigenstates are given by
\begin{equation*}
    \ket{s,\vec{k}} = \frac{1}{\sqrt{2}}\begin{pmatrix}
        1  \\
        -\ii s\exp{\ii s\phi_k}
    \end{pmatrix}.
\end{equation*}
In detail, we have
\begin{align*}
    \Braket{s,\vec{k}|\sigma^x|s',\vec{k'}} &= \frac{\ii}{2}\left[s\exp{-\ii \phi} - s'\exp{\ii\phi'}\right]  \\
    \Braket{s',\vec{k'}|\sigma^x|s,\vec{k}} &= \frac{\ii}{2}\left[s'\exp{-\ii\phi'} - s\exp{\ii\phi}\right] \\
    \Braket{s,\vec{k}|\sigma^y|s',\vec{k'}} &= -\frac{1}{2}\left[s\exp{-i\phi} + s'\exp{\ii\phi'}\right]  \\
    \Braket{s',\vec{k'}|\sigma^y|s,\vec{k}} &= -\frac{1}{2}\left[s\exp{\ii\phi} + s'\exp{-\ii\phi'}\right]  \\
    \Braket{s,\vec{k}|\sigma^z|s',\vec{k'}} &= \frac{1}{2}\left[1 - ss'\exp{\ii(\phi'-\phi)}\right]  \\
    \Braket{s',\vec{k'}|\sigma^z|s',\vec{k'}} &= \frac{1}{2}\left[1 - ss'\exp{\ii(\phi -\phi')}\right] ,
\end{align*}
where we defined $\phi = \phi_k$ and $\phi' = \phi_{k'}$
Therefore, the relevant products are: 
\begin{align*}
     \mathcal{M}^{xx}_{ss'}(\vec{k},\vec{k'}) &= \frac{1}{2}\left[1 - ss'\cos(\phi + \phi')\right] \\
     \mathcal{M}^{yy}_{ss'}(\vec{k},\vec{k'})&= \frac{1}{2}\left[1 + ss'\cos(\phi + \phi')\right] \\
    \mathcal{M}^{zz}_{ss'}(\vec{k},\vec{k'}) &= \frac{1}{2}\left[1 - ss'\cos(\phi - \phi')\right] \\
    \mathcal{M}^{xz}_{ss'}(\vec{k},\vec{k'}) &= \frac{\ii}{2}\left[s'\cos\phi' - s\cos\phi\right] = -\mathcal{M}^{zx}_{ss'}(\vec{k},\vec{k'})\\
    \mathcal{M}^{yz}_{ss'}(\vec{k},\vec{k'}) &= \frac{\ii}{2}\left[s\sin\phi - s'\sin\phi'\right] = -\mathcal{M}^{zy}_{ss'}(\vec{k},\vec{k'})
\end{align*}
We first consider the $i \neq j$ case, defining $\vec{r}_{ij} = -\vec{r}_{ji} = (R\cos\theta, R\sin\theta)$. The three types of integral which enter into the expression (\ref{eq:mspaceint}) are:
\begin{align*}
    \int\frac{\rmd^2\vec{q}}{(2\pi)^2}\exp{\pm\ii\vec{q}\cdot\vec{r}_{ij}}A_{s}(\vec{q},\omega) &= \frac{m}{2\pi}\int_{0}^\infty\rmd q \ q \delta(q - k_{0s}(\omega))\Theta(\omega - \epsilon_{0s})\int_0^{2\pi}\rmd\phi\exp{\pm\ii qR\cos(\phi-\theta)} \nonumber \\
    &= m\int_0^\infty\rmd q q \delta(q - k_{0s}(\omega))\Theta(\omega - \epsilon_{0s}) J_0(qR) \nonumber \\
    &= mJ_0(k_{0s}(\omega)R)\Theta(\omega - \epsilon_{0s}) \\
    \int\frac{\rmd^2\vec{q}}{(2\pi)^2}\exp{\pm\ii\vec{q}\cdot\vec{r}_{ij}}A_{s}(\vec{q},\omega)\cos\phi &= \pm \ii mJ_1(k_{0s}(\omega)R)\Theta(\omega - \epsilon_{0s})\cos\theta \\
     \int\frac{\rmd^2\vec{q}}{(2\pi)^2}\exp{\pm\ii\vec{q}\cdot\vec{r}_{ij}}A_{s}(\vec{q},\omega)\sin\phi &= \pm \ii mJ_1(k_{0s}(\omega)R)\Theta(\omega - \epsilon_{0s})\sin\theta 
\end{align*}
where $J_n$ is the Bessel function of the first kind at order $n$. Using the results above, we find after some algebra that the momentum space integrals have the following matrix structure:
\begin{align*}
    &\sum_{ss'}\int\frac{\rmd^2\vec{k}}{(2\pi)^2}\int\frac{\rmd^2\vec{k'}}{(2\pi)^2}\exp{-\ii(\vec{k} - \vec{k'})\cdot\vec{r}_{ij}}\mathcal{M}^{\mu\nu}_{ss'}(\vec{k},\vec{k'})A_{s}(\vec{k},\omega'')A_{s'}(\vec{k'},\omega') \\
    &= \frac{m^2}{2}\sum_{ss'}\frac{k_{0s}(\omega'')k_{0s'}(\omega')}{k_0(\omega')k_0(\omega'')}\mathds{J}^{\mu\nu}_{ss'}(\omega',\omega'';\vec{r}_{ij})\Theta(\omega'' - \epsilon_{0s})\Theta(\omega' - \epsilon_{0s'})
\end{align*}
which is Eq. (\ref{eq:mspaceintfull}) from the main text. \\
Now, let's consider the local case $i = j$. In this case, $\vec{r}_{ij} = \vec{0}$, which significantly simplifies the matrix structure of Eq. (\ref{eq:mspaceintfull}). We have:
\begin{align*}
    \int\frac{\rmd^2\vec{k}}{(2\pi)^2} A_s(\vec{k},\omega) &= 2\pi m\int_0^\infty\rmd k \ \frac{k}{k_0(\omega)} \delta(k - k_{0s}(\omega))\int_0^{2\pi}\rmd\phi \\
    &= m\int_0^\infty\rmd k \ \frac{k}{k_0(\omega)} \delta(k - k_{0s}(\omega)) \\
    &= \frac{mk_{0s}(\omega)}{k_0(\omega)}; \\
    \int\frac{\rmd^2\vec{k}}{(2\pi)^2} A_s(\vec{k},\omega)\cos\phi &= \frac{m}{2\pi} \int_0^\infty\rmd k \ \frac{k}{k_0(\omega)} \delta(k - k_{0s}(\omega))\int_0^{2\pi}\rmd\phi\cos\phi \\
    &= 0
    = \int\frac{\rmd^2\vec{k}}{(2\pi)^2} A_s(\vec{k},\omega)\sin\phi,
\end{align*}
making the structure of Eq. (\ref{eq:mspaceint}) much simpler. This is namely
\begin{align*}
    &\sum_{ss'}\int\frac{\rmd^2\vec{k}}{(2\pi)^2}\int\frac{\rmd^2\vec{k'}}{(2\pi)^2}\exp{-\ii(\vec{k} - \vec{k'})\cdot\vec{r}_{ij}}\mathcal{M}^{\mu\nu}_{ss'}(\vec{k},\vec{k'})A_{s}(\vec{k},\omega'')A_{s'}(\vec{k'},\omega')\\
    &= \left[\frac{m^2}{k_0(\omega')k_0(\omega'')}\sum_{ss'}k_{0s'}(\omega')k_{0s}(\omega'')\right]\mathds{1}
\end{align*}
\subsection{Computing the damping (Eq. (\ref{eq:alphatensor}))}
\subsubsection*{Non-local damping}
We may now proceed with the derivation of the non-local damping. We take Eq. (\ref{eq:retadfreqdom}) from the main text, and split the complex denominator into its real and imaginary parts according to the Sokhotski-Plemelj theorem, which states that:
\begin{equation*}
    \lim_{\epsilon\to 0}\frac{1}{x\pm \ii\epsilon} = \mathcal{P}\left(\frac{1}{x}\right) \mp\ii\pi\delta(x),
\end{equation*}
where $\mathcal{P}$ denotes the Cauchy principal value. For the delta function part, we therefore have:
\begin{align*}
    &\pm\frac{\ii\pi J_{sd}^2m^2}{4}\sum_{ss'}\int_{\epsilon_{0s'}}^\infty\frac{\rmd\omega'}{2\pi}\int_{\epsilon_{0s}}^\infty\frac{\rmd\omega''}{2\pi}\delta(\omega - (\omega'-\omega''))\left(N_\text{F}(\omega''-\mu) - N_\text{F}(\omega'-\mu)\right)\frac{k_{0s}(\omega'')k_{0s'}(\omega')}{k_0(\omega')k_0(\omega'')}\mathds{J}^{\mu\nu}_{ss'}(\omega',\omega'';\vec{r}_{ij})\nonumber \\
    &=\pm\frac{\ii\pi J_{sd}^2m^2}{4}\sum_{ss'}\int_{\epsilon_{0s'}}^\infty\frac{\rmd\omega'}{(2\pi)^2} \left(N_\text{F}(\omega'-(\omega + \mu)) - N_\text{F}(\omega'-\mu)\right)\frac{k_{0s}(\omega' - \omega)k_{0s'}(\omega')}{k_0(\omega')k_0(\omega' - \omega)}\mathds{J}^{\mu\nu}_{ss'}(\omega',\omega'-\omega;\vec{r}_{ij}) \nonumber \\
    &= \pm\frac{\ii\pi J_{sd}^2m^2}{4}\sum_{ss'}\int_{\epsilon_{0s'}}^\infty\frac{\rmd\omega'}{(2\pi)^2} \left(N_\text{F}(\omega'-(\omega + \mu)) - N_\text{F}(\omega'-\mu)\right)f(\omega, \omega') \quad \boxed{f(\omega, \omega') \equiv \frac{k_{0s}(\omega' - \omega)k_{0s'}(\omega')}{k_0(\omega')k_0(\omega' - \omega)}\mathds{J}^{\mu\nu}_{ss'}(\omega',\omega'-\omega;\vec{r}_{ij})}\nonumber \\
    &\approx\pm\frac{\ii J_{sd}^2m^2 \omega}{16\pi}\sum_{ss'}\int_{\epsilon_{0s'}}^\infty\rmd\omega'\frac{\partial}{\partial\omega}\left(N_\text{F}(\omega'-(\omega + \mu)) - N_\text{F}(\omega'-\mu)\right)f(\omega, \omega')\rvert_{\omega=0} \nonumber \\
    &= \pm\frac{\ii J_{sd}^2m^2 \omega}{16\pi}\sum_{ss'}\int_{\epsilon_{0s'}}^\infty\rmd\omega'f(\omega=0, \omega')\frac{\partial}{\partial\omega}\left(N_\text{F}(\omega'-(\omega + \mu)) - N_\text{F}(\omega'-\mu)\right) \nonumber \\
    &\approx \pm\frac{\ii J_{sd}^2m^2 \omega}{16\pi}\sum_{ss'}\int_{\epsilon_{0s'}}^\infty\rmd\omega'f(\omega=0, \omega')\delta(\omega'-\mu) \nonumber \\
    &\approx \pm\frac{\ii J_{sd}^2m^2 \omega}{16\pi}\sum_{ss'}\frac{k_{0s}(\mu)k_{0s'}(\mu)}{k_0(\mu)k_0(\mu)}\mathds{J}^{\mu\nu}_{ss'}(\mu,\mu;\vec{r}_{ij})
\end{align*}
which is identical to Eq. (\ref{eq:raimaglowfreq}). The use of the first delta function to collapse the $\omega''$ integral is always valid, whereas the use of the second delta function (which comes from approximating the Fermi function's derivative as a delta function provided $T\ll T_F$) to collapse the $\omega'$ integral is valid provided $\mu > 0$. So, the non-local damping is found from Eq. (\ref{eq:dampingident}) to be
\begin{align*}
    \vec{\alpha}_{ij} &= 2\lim_{\omega\to 0}\frac{1}{\omega}\frac{J_{sd}^2m^2 \omega}{16\pi}\sum_{ss'}\frac{k_{0s}(\mu)k_{0s'}(\mu)}{k_0(\mu)k_0(\mu)}\mathds{J}^{\mu\nu}_{ss'}(\mu,\mu;\vec{r}_{ij}) \nonumber \\
    &=\frac{J_{sd}^2m^2}{8\pi}\sum_{ss'}\frac{k_{0s}(\mu)k_{0s'}(\mu)}{k_0(\mu)k_0(\mu)}\mathds{J}^{\mu\nu}_{ss'}(\mu,\mu;\vec{r}_{ij})
\end{align*}
To make connection with Eq. (\ref{eq:dampspec}) from the main text, we consider the case where $\theta = 0$. One finds in this case:
\begin{align*}
    \vec{\alpha}^{xx}_{ij} &= \frac{\alpha}{4}\sum_{ss'}\left[J_0(k_{Fs}R)J_0(k_{Fs'}R) - ss'J_1(k_{Fs}R)J_1(k_{Fs'}R)\right] = \vec{\alpha}^{zz}_{ij} \equiv \alpha_0 \\
    \vec{\alpha}^{yy}_{ij} &= \frac{\alpha}{4}\sum_{ss'}\left[J_0(k_{Fs}R)J_0(k_{Fs'}R) - ss'J_1(k_{Fs}R)J_1(k_{Fs'}R)\right]  \equiv \alpha_2 \\
    \vec{\alpha}^{xz}_{ij} &= -\frac{\alpha}{4}\sum_{ss'}\left(s'J_1(k_{Fs'}RJ_0(k_{Fs}R)+sJ_0(k_{Fs'}R)J_1(k_{Fs}R)\right) \equiv -\alpha_1 = -\vec{\alpha}^{zx}_{ij} \\
    \vec{\alpha}^{yz}_{ij} &= \vec{\alpha}^{zy}_{ij} = \vec{\alpha}^{xy}_{ij} = \vec{\alpha}^{yx}_{ij} = 0
\end{align*}
\subsubsection*{Local damping}
We can follow an identical derivation as was used above to compute the local damping. The result is
\begin{equation*}
    \alpha = 2\frac{J_{sd}^2m^2}{16\pi}\sum_{ss'}\frac{k_{0s}(\mu)k_{0s'}(\mu)}{k_0(\mu)k_0(\mu)} = \frac{J_{sd}^2m^2}{2\pi}
\end{equation*}
\subsection{Computing the exchange (Eq. (\ref{eq:Ktensor}))}
We begin with Eq. (\ref{eq:pirareal}), having substituted Eq. (\ref{eq:mspaceintfull})
\begin{align*}
    \lim_{\omega\to 0}\Re\left[\left(\vec{\Pi}^{R(A)}\right)^{\mu\nu}_{ij}(\omega)\right] &= \lim_{\omega\to 0}-\frac{J_{sd}^2m^2}{4}\sum_{ss'}\mathcal{P}\int\frac{\rmd\omega'}{2\pi}\int\frac{\rmd\omega''}{2\pi}\frac{1}{\omega - (\omega'-\omega'')}\left(N_\text{F}(\omega''-\mu) - N_\text{F}(\omega'-\mu)\right) \nonumber \\
    &\times\frac{k_{0s}(\omega'')k_{0s'}(\omega')}{k_0(\omega')k_0(\omega'')}\mathds{J}^{\mu\nu}_{ss'}(\omega',\omega'';\vec{r}_{ij})\Theta(\omega'' - \epsilon_{0s})\Theta(\omega' - \epsilon_{0s'}).
\end{align*}
Now define
\begin{equation*}
    \mathcal{F}^{\mu\nu}_{ss'}(\omega',\omega'') = \frac{k_{0s}(\omega'')k_{0s'}(\omega')}{k_0(\omega')k_0(\omega'')}\mathds{J}^{\mu\nu}_{ss'}(\omega',\omega'';\vec{r}_{ij})\Theta(\omega'' - \epsilon_{0s})\Theta(\omega' - \epsilon_{0s'}).
\end{equation*}
One observes that $\mathcal{F}$ has the property
\begin{equation*}
    \mathcal{F}^{\mu\nu}_{ss'}(\omega',\omega'') = \mathcal{F}^{\mu\nu}_{s's}(\omega'',\omega')
\end{equation*}
under the involution
\begin{equation*}
    (\omega'', s) \longleftrightarrow (\omega',s')
\end{equation*}
for all $\mu\nu$. On the other hand, the difference of Fermi functions changes sign under such an involution, and the denominator is also transformed to $1/(\omega - \omega'' + \omega')$ Thus we may replace the integral above with its symmetrised version (up to a prefactor $-J_{sd}^2m^2/4$):
\begin{align*}
     \lim_{\omega\to 0}\Re\left[\left(\vec{\Pi}^{R(A)}\right)^{\mu\nu}_{ij}(\omega)\right] &= \frac{1}{2}\lim_{\omega\to 0}\sum_{ss'}\mathcal{P}\int\frac{\rmd\omega'}{2\pi}\int\frac{\rmd\omega''}{2\pi}\left(N_\text{F}(\omega''-\mu) - N_\text{F}(\omega'-\mu)\right)\mathcal{F}^{\mu\nu}_{ss'}(\omega',\omega'')\left[\frac{1}{\omega - \Delta} - \frac{1}{\omega + \Delta}\right] \nonumber \\
     &= \lim_{\omega\to 0}\sum_{ss'}\mathcal{P}\int\frac{\rmd\omega'}{2\pi}\int\frac{\rmd\omega''}{2\pi}\left(N_\text{F}(\omega''-\mu) - N_\text{F}(\omega'-\mu)\right)\mathcal{F}^{\mu\nu}_{ss'}(\omega',\omega'')\frac{\omega'-\omega''}{\omega^2 - (\omega'-\omega'')^2} \\
\end{align*}
We may now replace the Fermi functions with step functions. For generic $s,s'$, the integrand gives two contributions: $+1$ for $\infty > \omega' > \mu$ and $\mu > \omega'' > \epsilon_{0s}$ (, and $-1$ for $\infty > \omega'' > \mu$ and $\mu > \omega' > \epsilon_{0s'}$. The integral becomes:
\begin{align*}
    &\lim_{\omega\to 0}\sum_{ss'}\mathcal{P}\int_\mu^\infty\frac{\rmd\omega'}{2\pi}\int_{\epsilon_{0s}}^\mu\frac{\rmd\omega''}{2\pi}\mathcal{F}^{\mu\nu}_{ss'}(\omega',\omega'')G(\omega, \Delta)\\
    &-\lim_{\omega\to 0}\sum_{ss'}\mathcal{P}\int_{\epsilon_{0s'}}^\mu\frac{\rmd\omega'}{2\pi}\int_{\mu}^\infty\frac{\rmd\omega''}{2\pi}\mathcal{F}^{\mu\nu}_{ss'}(\omega',\omega'')G(\omega, \Delta) \\
    &=\lim_{\omega\to 0}\sum_{ss'}\mathcal{P}\frac{1}{(2\pi)^2}\int\int_A \rmd\omega'\rmd\omega''\mathcal{F}^{\mu\nu}_{ss'}(\omega',\omega'')G(\omega, \Delta) \\
    &-\lim_{\omega\to 0}\sum_{ss'}\mathcal{P}\frac{1}{(2\pi)^2}\int\int_B \rmd\omega'\rmd\omega''\mathcal{F}^{\mu\nu}_{ss'}(\omega',\omega'')G(\omega, \Delta)
\end{align*}
where we defined
\begin{equation*}
    G(\omega, \Delta) = \frac{\Delta}{\omega^2 - \Delta^2} = -G(\omega, -\Delta)
\end{equation*}
In the second term, we then perform the same involution as defined above, which maps the domain of integration $B$ to the domain of integration $A$ in the second term, while also introducing an extra minus sign thanks to the structure of $G$. This involution lets us combine the two integrals to obtain
\begin{align*}
    &2\lim_{\omega\to 0}\sum_{ss'}\mathcal{P}\frac{1}{(2\pi)^2}\int\int_A \rmd\omega'\omega''\mathcal{F}^{\mu\nu}_{ss'}(\omega',\omega'')G(\omega, \Delta) \\
    &= 2\sum_{ss'}\int_\mu^\infty\frac{\rmd\omega'}{2\pi}\int_{\epsilon_{0s}}^\mu\frac{\rmd\omega''}{2\pi}\frac{k_{0s}(\omega'')k_{0s'}(\omega')}{k_0(\omega')k_0(\omega'')}\mathds{J}^{\mu\nu}_{ss'}(\omega',\omega'';\vec{r}_{ij})\lim_{\omega\to 0}\frac{\Delta}{\omega^2 -\Delta^2} \\
    &= -2\sum_{ss'}\int_\mu^\infty\frac{\rmd\omega'}{2\pi}\int_{\epsilon_{0s}}^\mu\frac{\rmd\omega''}{2\pi}\frac{1}{\omega'-\omega''}\frac{k_{0s}(\omega'')k_{0s'}(\omega')}{k_0(\omega')k_0(\omega'')}\mathds{J}^{\mu\nu}_{ss'}(\omega',\omega'';\vec{r}_{ij}),
\end{align*}
and thus, combining all prefactors, we see that the integrand always has the form
\begin{equation*}
    \frac{J_{sd}^2m^2}{8\pi^2}\sum_{ss'}\int_\mu^\infty\rmd\omega'\int_{\epsilon_{0s}}^\mu\rmd\omega''\frac{1}{\omega'-\omega''}\frac{k_{0s}(\omega'')k_{0s'}(\omega')}{k_0(\omega')k_0(\omega'')}\mathds{J}^{\mu\nu}_{ss'}(\omega',\omega'';\vec{r}_{ij})
\end{equation*}
We can now make some approximations. Firstly, observing the denominator, we see that the divergent behaviour has gone from a line singularity to a single divergent point at $\omega' = \omega'' = \mu$, making the application of the principal value again valid. Observing the structure of the integrand, one notes that the Bessel functions are all decaying functions of magnitude less than or equal to $1$, whereas the factor
\begin{equation*}
    \left(1 - \frac{sq_R}{k_0(\omega'')}\right)\left(1 - \frac{s'q_R}{k_0(\omega')}\right),
\end{equation*}
is finite and also of order $1$ at the pole, whereas the divergence of this factor as $\omega'\to\infty$ is controlled by the denominator $1/(\omega'-\omega'')$. Hence the pole gives the largest contribution to the integrand, which motivates the expansion of the $k_0$ factors around this point. We know that 
\begin{align*}
    &k_{0s}(\omega) = -sq_R + \sqrt{2m\omega + q_R^2} \\
    \implies & \frac{\partial k_{0s}(\omega)}{\partial\omega}\bigg\rvert_{\omega=\mu} = \frac{m}{\sqrt{2m\mu + q_R^2}} = \frac{m}{k_F} \\
    \implies & k_{0s}(\omega) \approx k_F - sq_R + \frac{m}{k_F}\left(\omega-\mu\right)
\end{align*}
The prefactor 
\begin{equation*}
    \left(1 - \frac{sq_R}{k_0(\omega'')}\right)\left(1 - \frac{s'q_R}{k_0(\omega')}\right),
\end{equation*}
may be replaced with
\begin{equation*}
    \left(1 - \frac{sq_R}{k_F}\right)\left(1 - \frac{s'q_R}{k_F}\right).
\end{equation*}
which is $1$ to leading order if we also make the assumption that $q_R \ll k_F$, similar to how is done in Ref. \cite{Imamura_PhysRevB_69_121303_2004}. \\ 
With this expansion in mind, we are ready to apply a further approximation. The Bessel functions are decaying oscillatory functions. We may apply an asymptotic limit to these functions, because this asymptotic behaviour takes over quite quickly when the argument is relatively small. Then, a product of the form $J_v(k_{0s}(\omega'')R)J_{\nu'}(k_{0s'}(\omega')R)$ (of which all the components of Eq. (\ref{eq:mspaceintfull}) are linear combinations) has the asymptotic form:
\begin{align*}
    &\frac{2}{\pi R\sqrt{kk'}}\cos\left(k_{0s}(\omega'')R - \frac{(\pi + 2\nu)}{4}\right)\cos\left(k_{0s}(\omega')R - \frac{(\pi + 2\nu')}{4}\right)  \\
    &\approx \frac{2}{\pi k_F R}\cos\left(\left(k_F - sq_R + \frac{\omega'' - \mu}{v_F}\right)R - \phi_\nu\right)\cos\left(\left(k_F - s'q_R + \frac{\omega' - \mu}{v_F}\right)R - \phi_{\nu'}\right) \label{eq:bessasymp} \boxed{\Omega' = \omega'-\mu; \ \Omega'' = \omega'' - \mu} \nonumber \\
    &= \frac{2}{\pi k_F R}\cos\left(\left(k_F - sq_R + \frac{\Omega''}{v_F}\right)R - \phi_\nu\right)\cos\left(\left(k_F - s'q_R + \frac{\Omega'}{v_F}\right)R  - \phi_{\nu'}\right) \boxed{\cos A\cos B = \frac{1}{2}\left[\cos(A+B)+ \cos(A-B)\right]} \\
    &=\frac{1}{\pi k_FR}\cos\left(\left(\frac{\Omega'' + \Omega'}{v_F} + \Phi^{ss'}_{\nu\nu',+}\right)R\right) \boxed{\Phi^{ss'}_{\nu\nu', +} = 2k_F - (s+s')q_R -(\phi_{\nu} + \phi_{\nu'})/R}\nonumber \\
    &+ \frac{1}{\pi k_FR}\cos\left(\left(\frac{\Omega'' - \Omega'}{v_F} + \Phi^{ss'}_{\nu\nu',-}\right)R\right) \boxed{\Phi^{ss'}_{\nu\nu', -} = (s'-s)q_R -((\phi_{\nu} - \phi_{\nu'})/R}
\end{align*}
where $\phi_\nu = (\pi + 2\nu)/4$ (and similarly for $\phi_{\nu'}$). Now, let's consider plugging each of these terms into the integral. For the sum term, noting that we already made the substitutions $\Omega' = \omega' -\mu$, $\Omega'' = \omega'' - \mu$, and remembering an overall prefactor of:
\begin{equation*}
    \frac{J_{sd}^2m^2}{8\pi^2}\frac{1}{\pi k_FR}
\end{equation*}
we must have an integral of the form
\begin{align*}
    \mathcal{P}\int_0^\infty\rmd\Omega'\int_{\epsilon_{0s} -\mu}^0\rmd\Omega''\frac{\cos\left(\left(\frac{\Omega'' + \Omega'}{v_F} + \Phi^{ss'}_{\nu\nu',+}\right)R\right)}{\Omega'-\Omega''}
\end{align*}
We now define some variables for conciseness' sake. Let $A = R\Phi^{ss'}_{\nu\nu',+}$ and $\Gamma = R/v_F$. Then: 
\begin{align*}
     \int_0^\infty\rmd\Omega'\int_{\epsilon_{0s} -\mu}^0\rmd\Omega''\frac{\cos\left(\Gamma(\Omega'' + \Omega') + A\right)}{\Omega'-\Omega''} &\approx \int_0^\infty\rmd\Omega'\int_{-\infty}^0\rmd\Omega''\frac{\cos\left(\Gamma(\Omega'' + \Omega') + A\right)}{\Omega'-\Omega''}.
\end{align*}
where the lower bound of the $\Omega''$ integral may now be extended to $-\infty$, because contributions in this region are suppressed by the denominator. The structure of the integrand suggests the substitutions $\Omega' - \Omega'' = v$ and $\Omega' + \Omega'' = u$. These transform the integral to
\begin{align*}
    \int_0^\infty\rmd\Omega'\int_{-\infty}^0\rmd\Omega''\frac{\cos\left(\Gamma(\Omega'' + \Omega') + A\right)}{\Omega'-\Omega''} &= \frac{1}{2}\int_0^\infty\frac{\rmd v}{v}\int_{-v}^{v}\rmd u \cos(\Gamma u + A) \\
    &= \frac{1}{\Gamma}\int_0^\infty\frac{\rmd v}{v}\left[\sin(\Gamma u + A)\right]^{v}_{-v} \\
    &=  \frac{1}{\Gamma}\int_0^\infty\frac{\rmd v}{v}\left[\sin(\Gamma v + A) - \sin(-\Gamma v + A)\right] \\
    &= \frac{1}{\Gamma}\int_0^\infty\frac{\rmd v}{v}\cos(A)\sin(v) \\
    &= \frac{\pi}{2\Gamma}
\end{align*}
using the well-known result for the Dirichlet integral. \\
The final steps are simply to consider the four cases for the phase $\phi_{\nu} + \phi_{\nu'}$, and plug the results into the sum over $s$ and $s'$. If $\nu = \nu' = 0$, this is $\pi/2$. If $\nu = 0$ and $\nu' = 1$ (or vice versa) this is $\pi$. If $\nu = \nu' = 1$, this is $3\pi/2$. We thus find the following phase shifts:
\begin{equation*}
\begin{array}{c|c}
    \nu, \nu' &  \cos\left(\Phi^{\nu\nu'}_{ss'}R\right) \\
     0, 0 &  \sin\left(2k_FR - (s+s')q_RR\right) \\
     0, 1 & -\cos\left(2k_FR - (s+s')q_RR\right) \\
     1, 1 & -\sin\left(2k_FR - (s+s')q_RR\right)
\end{array}
\end{equation*}
Restoring all prefactors, we have an overall prefactor for each term of
\begin{equation*}
    \frac{J_{sd}^2m^2}{8\pi^2}\frac{1}{\pi k_FR}\frac{\pi}{2\Gamma} = \frac{J_{sd}^2 m k_F^2}{16\pi^2}\frac{1}{(k_FR)^2}
\end{equation*}
The summations over helicities $s, \ s' = \pm 1$ are easily evaluated for each component, and are summarised below:
\begin{align*}
    &\sum_{ss'}\sin(2k_FR - (s+s')q_RR)\left(1 + ss'\cos(2\theta)\right) \quad \boxed{xx}\\
    &= 2\sin(2k_FR)(1 - \cos(2\theta)) \\
    &+ \sin(2k_FR - 2q_RR)(1 + \cos(2\theta)) \\
    &+ \sin(2k_FR + 2q_RR)(1 + \cos(2\theta)) \\
    &= 2\sin(2k_FR)\left[1 + \cos(2q_RR) - \cos(2\theta)(1 - \cos(2q_RR))\right]  \\
    &\sum_{ss'}\sin(2k_FR - (s+s')q_RR)\left(1 - ss'\cos(2\theta)\right) \quad \boxed{yy}\\
    &= 2\sin(2k_FR)(1 + \cos(2\theta)) \\
    &+ (\sin(2k_FR - 2q_RR) + \sin(2k_FR + 2q_RR))(1 - \cos(2\theta)) \\
    &= 2\sin(2k_FR)\left[1 + \cos(2q_RR) + \cos(2\theta)(1 - \cos(2q_RR))\right]  \\
    &\sum_{ss'}\sin(2k_FR - (s+s')q_RR)\left(1 + ss'\right) \quad \boxed{zz}\\
    &= 4\sin(2k_FR)\cos(2q_RR)  \\
    &+\cos\theta\sum_{ss'}\cos(2k_FR - (s+s')q_RR)(s' + s) \quad \boxed{xz}\\
    &= \cos\theta\left[\cos(2k_FR - 2q_RR) -\cos(2k_FR + 2q_RR)\right] \\
    &= 2\cos\theta\sin(2k_FR)\sin(2q_RR)  \\
    & -2\sin\theta\sin(2k_FR)\sin(2q_RR)  \quad\boxed{yz}\\
    &\frac{1}{2}\sin(2\theta)\sum_{ss'}ss'\sin(2k_FR - (s+s')q_RR) \\
    &= \frac{1}{2}\sin(2\theta)\left[\sin(2k_FR - 2q_RR) + \sin(2k_FR + 2q_RR) - 2\sin(2k_FR)\right] \\
    &= \sin(2\theta)\sin(2k_FR)\left[\cos(2q_RR) - 1\right] \quad \boxed{xy}\\
\end{align*}
which allows us to give a final expression for the exchange tensor, using Eq. (\ref{eq:fieldident}):
\begin{align*}
    \vec{K}_{ij} = 2\lim_{\omega\to 0}\Re\left[\left(\vec{\Pi}^R\right)^{\mu\nu}_{ij}(\omega)\right] &= \frac{J_{sd}^2mk_F^2}{4\pi^2}\frac{\sin(2k_FR)}{(k_FR)^2}\mathds{M}(q_RR;\theta) \\
    &= KF(k_FR)\mathds{M}(q_RR;\theta)
\end{align*}
where
\begin{align*}
    K &= \frac{J_{sd}^2mk_F^2}{4\pi^2}; \\
    F(k_FR) &= \frac{\sin(2k_FR)}{(k_FR)^2}; \\
    \mathds{M}(q_RR;\theta) &= \begin{pmatrix}
        1 + \cos(2q_RR) - \cos(2\theta)(1-\cos(2q_RR))& \frac{1}{2}\sin(2\theta)(\cos(2q_RR) - 1) & \cos\theta\sin(2q_RR) \\
        \frac{1}{2}\sin(2\theta)(\cos(2q_RR) - 1) & 1 + \cos(2q_RR) + \cos(2\theta)(1-\cos(2q_RR)) & -\sin\theta\sin(2q_RR) \\ 
        - \cos\theta\sin(2q_RR) & \sin\theta\sin(2q_RR) & 2\cos(2q_RR),
    \end{pmatrix}
\end{align*}
exactly as given in the main text. In Sec. \ref{sec:llg} of the main text, we also specialise to the case where $\theta=0$; to this end, we quote the matrix $\mathds{M}(q_RR;\theta=0)$ below and identify expressions for the exchange constants $J_0$, $J_1$, $D$ given in Eq. (\ref{eq:exchnspec}):
\begin{equation*}
    \mathds{M}(q_RR;\theta) = \begin{pmatrix}
        2\cos(2q_RR)& 0 & \sin(2q_RR) \\
        0 & 2  & 0\\ 
        -\sin(2q_RR) & 0 & 2\cos(2q_RR),
    \end{pmatrix},
\end{equation*}
allowing us to identify:
\begin{align*}
    \frac{J_0}{KF(k_FR)} &= 2\cos(2q_RR) \\
    \frac{J_1}{KF(k_FR)} &= 2 \\
    \frac{D}{KF(k_FR)} &= \sin(2q_RR)
\end{align*}
\end{widetext}
\bibliography{biblio}

@article{Yuan_PhysRevB_107_024418_2023,
  author    = {Yuan, H. Y. and Lavrijsen, R. and Duine, R. A.},
  title     = {{Unidirectional} magnetic coupling induced by chiral interaction and nonlocal damping},
  journal   = {Phys. Rev. B},
  issue     = {2},
  volume    = {107},
  pages     = {024418},
  year      = {2023},
  month     = {1},
  publisher = {American Physical Society (APS)},
  doi       = {10.1103/physrevb.107.024418},
  url       = {https://dx.doi.org/10.1103/physrevb.107.024418}
}

@article{Zou_PhysRevLett_132_036701_2024,
  author    = {Zou, Ji and Bosco, Stefano and Thingstad, Even and Klinovaja, Jelena and Loss, Daniel},
  title     = {{Dissipative} {Spin}-{Wave} {Diode} and {Nonreciprocal} {Magnonic} {Amplifier}},
  journal   = {Phys. Rev. Lett.},
  issue     = {3},
  volume    = {132},
  pages     = {036701},
  year      = {2024},
  month     = {1},
  publisher = {American Physical Society (APS)},
  doi       = {10.1103/physrevlett.132.036701},
  url       = {https://dx.doi.org/10.1103/physrevlett.132.036701}
}

@book{Sze_SemiconductorDevices,
  author    = {S. M. Sze and Kwok K. Ng},
  title     = {Physics of Semiconductor Devices},
  edition   = {3rd},
  year      = {2006},
  publisher = {Wiley-Interscience},
  address   = {Hoboken, NJ},
  isbn      = {978-0-471-14323-9},
  doi       = {10.1002/0470068329}
}

@article{Chumak_MagnonSpintronics,
  author  = {Chumak, A. V. and Vasyuchka, V. I. and Serga, A. A. and Hillebrands, B.},
  title   = {Magnon spintronics},
  journal = {Nature Physics},
  year    = {2015},
  volume  = {11},
  pages   = {453--461},
  doi     = {10.1038/nphys3347}
}

@article{Chumak_MagnonTransistor,
  author  = {Chumak, A. V. and Serga, A. A. and Hillebrands, B.},
  title   = {Magnon transistor for all-magnon data processing},
  journal = {Nature Communications},
  year    = {2014},
  volume  = {5},
  pages   = {4700},
  doi     = {10.1038/ncomms5700}
}

@article{Lenk_BuildingBlocksMagnonics,
  author  = {Lenk, B. and Ulrichs, H. and Garbs, F. and M{\"u}nzenberg, M.},
  title   = {The building blocks of magnonics},
  journal = {Physics Reports},
  year    = {2011},
  volume  = {507},
  number  = {4--5},
  pages   = {107--136},
  doi     = {10.1016/j.physrep.2011.06.003}
}

@article{Lan_SpinWaveDiode,
  author  = {Lan, J. and Yu, W. and Wu, R. and Xiao, J.},
  title   = {Spin-Wave Diode},
  journal = {Physical Review X},
  year    = {2015},
  volume  = {5},
  number  = {4},
  pages   = {041049},
  doi     = {10.1103/PhysRevX.5.041049}
}

@article{DamonEshbach_MagnetostaticModes,
  author  = {Damon, R. W. and Eshbach, J. R.},
  title   = {Magnetostatic modes of a ferromagnet slab},
  journal = {Journal of Physics and Chemistry of Solids},
  year    = {1961},
  volume  = {19},
  number  = {3--4},
  pages   = {308--320},
  doi     = {10.1016/0022-3697(61)90041-5}
}

@article{Barman_MagnonicsRoadmap2021,
  author    = {Barman, Anjan and Gubbiotti, Gianluca and Ladak, S. and Adeyeye, A. O. and Krawczyk, M. and Gr{\"a}fe, J. and Adelmann, C. and Cotofana, S. and Naeemi, A. and Vasyuchka, V. I. and Hillebrands, B. and Nikitov, S. A. and Yu, H. and Grundler, D. and Sadovnikov, A. V. and Grachev, A. A. and Sheshukova, S. E. and Duquesne, J.-Y. and Marangolo, M. and Csaba, G. and Porod, W. and Demidov, V. E. and Urazhdin, S. and Demokritov, S. O. and Albisetti, E. and Petti, D. and Bertacco, R. and Schultheiss, H. and Kruglyak, V. V. and Poimanov, V. D. and Sahoo, S. and Sinha, J. and Yang, H. and Münzenberg, M. and Moriyama, T. and Mizukami, S. and Landeros, P. and Gallardo, R. A. and Carlotti, G. and Kim, J.-V. and Stamps, R. L. and Camley, R. E. and Rana, B. and Otani, Y. and Yu, W. and Yu, T. and Bauer, G. E. W. and Back, C. and Uhrig, G. S. and Dobrovolskiy, O. V. and Budinska, B. and Qin, H. and van Dijken, S. and Chumak, A. V. and Khitun, A. and Nikonov, D. E. and Young, I. A. and Zingsem, B. W. and Winklhofer, M.},
  title     = {The 2021 {M}agnonics {R}oadmap},
  journal   = {Journal of Physics: Condensed Matter},
  year      = {2021},
  volume    = {33},
  number    = {41},
  pages     = {413001},
  doi       = {10.1088/1361-648X/abec1a}
}

@article{Flebus_MagnonicsRoadmap2024,
  author    = {Flebus, Benedetta and Grundler, Dirk and Rana, Bivas and Otani, Yoshi Chika and Barsukov, Igor and Gubbiotti, Gianluca and Landeros, Pedro and Wang, Qi and van der Sar, Toeno and {et al.}},
  title     = {The 2024 {M}agnonics {R}oadmap},
  journal   = {Journal of Physics: Condensed Matter},
  year      = {2024},
  volume    = {36},
  number    = {36},
  pages     = {363501},
  doi       = {10.1088/1361-648X/ad399c}
}

@article{Shockley_TheoryPNJunctions,
  author  = {Shockley, William},
  title   = {The Theory of p-n Junctions in Semiconductors and p-n Junction Transistors},
  journal = {Bell System Technical Journal},
  year    = {1949},
  volume  = {28},
  number  = {3},
  pages   = {435--489},
  doi     = {10.1002/j.1538-7305.1949.tb03645.x}
}

@book{BreuerPetruccione_OpenQuantumSystems,
  author    = {Breuer, Heinz-Peter and Petruccione, Francesco},
  title     = {The Theory of Open Quantum Systems},
  year      = {2002},
  publisher = {Oxford University Press},
  address   = {Oxford},
  isbn      = {978-0-19-852063-4}
}

@book{Yosida_TheoryMagnetism,
  author    = {Yosida, Kei},
  title     = {Theory of Magnetism},
  publisher = {Springer},
  series    = {Springer Series in Solid-State Sciences},
  volume    = {122},
  year      = {1996},
  address   = {Berlin, Heidelberg},
  isbn      = {978-3-540-60260-9},
  doi       = {10.1007/978-3-642-60937-7}
}

@book{AltlandSimons_CondMatFieldTheory,
  author    = {Altland, Alexander and Simons, Ben},
  title     = {Condensed Matter Field Theory},
  edition   = {2nd},
  year      = {2010},
  publisher = {Cambridge University Press},
  address   = {Cambridge},
  isbn      = {978-0-521-76001-7},
  doi       = {10.1017/CBO9780511779984}
}

@article{Shnirman_PhysRevLett_114_176806_2015,
  author    = {Shnirman, Alexander and Gefen, Yuval and Saha, Arijit and Burmistrov, Igor S. and Kiselev, Mikhail N. and Altland, Alexander},
  title     = {{Geometric} {Quantum} {Noise} of {Spin}},
  journal   = {Phys. Rev. Lett.},
  issue     = {17},
  volume    = {114},
  pages     = {176806},
  year      = {2015},
  month     = {4},
  publisher = {American Physical Society (APS)},
  doi       = {10.1103/physrevlett.114.176806},
  url       = {https://dx.doi.org/10.1103/physrevlett.114.176806}
}

@article{Verstraten_PhysRevRes_5_033128_2023,
  author    = {Verstraten, R. C. and Ludwig, T. and Duine, R. A. and Morais Smith, C.},
  title     = {{Fractional} {Landau}-{Lifshitz}-{Gilbert} equation},
  journal   = {Phys. Rev. Res.},
  issue     = {3},
  volume    = {5},
  pages     = {033128},
  year      = {2023},
  month     = {8},
  publisher = {American Physical Society (APS)},
  doi       = {10.1103/physrevresearch.5.033128},
  url       = {https://dx.doi.org/10.1103/physrevresearch.5.033128}
}

@book{Kamenev_FieldTheoryNonEq,
  author    = {Kamenev, Alex},
  title     = {Field Theory of Non-Equilibrium Systems},
  year      = {2011},
  publisher = {Cambridge University Press},
  address   = {Cambridge},
  isbn      = {978-0-521-76654-3},
  doi       = {10.1017/CBO9780511973445}
}

@article{Quarenta_PhysRevLett_133_136701_2024,
  author    = {Quarenta, Mario Gaspar and Tharmalingam, Mithuss and Ludwig, Tim and Yuan, H. Y. and Karwacki, Lukasz and Verstraten, Robin C. and Duine, Rembert A.},
  title     = {{Bath}-{Induced} {Spin} {Inertia}},
  journal   = {Phys. Rev. Lett.},
  issue     = {13},
  volume    = {133},
  pages     = {136701},
  year      = {2024},
  month     = {9},
  publisher = {American Physical Society (APS)},
  doi       = {10.1103/physrevlett.133.136701},
  url       = {https://dx.doi.org/10.1103/physrevlett.133.136701}
}

@article{Nunez_PhysRevB_77_054401_2008,
  author    = {N{\'u}{\~n}ez, Alvaro S. and Duine, R. A.},
  title     = {{Effective} temperature and {Gilbert} damping of a current-driven localized spin},
  journal   = {Phys. Rev. B},
  issue     = {5},
  volume    = {77},
  pages     = {054401},
  year      = {2008},
  month     = {2},
  publisher = {American Physical Society (APS)},
  doi       = {10.1103/physrevb.77.054401},
  url       = {https://dx.doi.org/10.1103/physrevb.77.054401}
}

@article{Duine_PhysRevB_75_214420_2007,
  author    = {Duine, R. A. and N{\'u}{\~n}ez, A. S. and Sinova, Jairo and MacDonald, A. H.},
  title     = {{Functional} {Keldysh} theory of spin torques},
  journal   = {Phys. Rev. B},
  issue     = {21},
  volume    = {75},
  pages     = {214420},
  year      = {2007},
  month     = {6},
  publisher = {American Physical Society (APS)},
  doi       = {10.1103/physrevb.75.214420},
  url       = {https://dx.doi.org/10.1103/physrevb.75.214420}
}

@book{Stoof_QuantumFluids,
  author    = {Stoof, H. T. C. and Gubbels, K. and Dickerscheid, D.},
  title     = {Ultracold Quantum Fields},
  year      = {2009},
  publisher = {Springer},
  address   = {Dordrecht},
  isbn      = {978-1-4020-9634-1},
  doi       = {10.1007/978-1-4020-9635-8}
}

@article{Hubbard_PhysRevLett_3_77_1959,
  author    = {Hubbard, J.},
  title     = {{Calculation} of {Partition} {Functions}},
  journal   = {Phys. Rev. Lett.},
  issue     = {2},
  volume    = {3},
  pages     = {77--78},
  year      = {1959},
  month     = {7},
  publisher = {American Physical Society (APS)},
  doi       = {10.1103/physrevlett.3.77},
  url       = {https://dx.doi.org/10.1103/physrevlett.3.77}
}

@article{Imamura_PhysRevB_69_121303_2004,
  author    = {Imamura, Hiroshi and Bruno, Patrick and Utsumi, Yasuhiro},
  title     = {{Twisted} exchange interaction between localized spins embedded in a one- or two-dimensional electron gas with {Rashba} spin-orbit coupling},
  journal   = {Phys. Rev. B},
  issue     = {12},
  volume    = {69},
  pages     = {121303},
  year      = {2004},
  month     = {3},
  publisher = {American Physical Society (APS)},
  doi       = {10.1103/physrevb.69.121303},
  url       = {https://dx.doi.org/10.1103/physrevb.69.121303}
}

@article{Turco_PhysRevRes_6_L022061_2024,
  author    = {Turco, Elia and Aapro, Markus and Ganguli, Somesh C. and Krane, Nils and Drost, Robert and Sobrino, Nahual and Bernhardt, Annika and Jur{\'\i}{\v{c}}ek, Michal and Fasel, Roman and Ruffieux, Pascal and Liljeroth, Peter and Jacob, David},
  title     = {{Demonstrating} {Kondo} behavior by temperature-dependent scanning tunneling spectroscopy},
  journal   = {Phys. Rev. Res.},
  issue     = {2},
  volume    = {6},
  pages     = {L022061},
  year      = {2024},
  month     = {6},
  publisher = {American Physical Society (APS)},
  doi       = {10.1103/physrevresearch.6.l022061},
  url       = {https://dx.doi.org/10.1103/physrevresearch.6.l022061}
}

@article{Zhang_NatCommun_4_2110_2013,
  author    = {Zhang, Yong-hui and Kahle, Steffen and Herden, Tobias and Stroh, Christophe and Mayor, Marcel and Schlickum, Uta and Ternes, Markus and Wahl, Peter and Kern, Klaus},
  title     = {{Temperature} and magnetic field dependence of a {Kondo} system in the weak coupling regime},
  journal   = {Nat. Commun.},
  issue     = {1},
  volume    = {4},
  pages     = {2110},
  year      = {2013},
  month     = {7},
  publisher = {Springer Science and Business Media LLC},
  doi       = {10.1038/ncomms3110},
  url       = {https://dx.doi.org/10.1038/ncomms3110}
}

@article{Madhavan_PhysRevB_64_165412_2001,
  author    = {Madhavan, V. and Chen, W. and Jamneala, T. and Crommie, M. F. and Wingreen, Ned S.},
  title     = {{Local} spectroscopy of a {Kondo} impurity: {Co} on {Au}(111)},
  journal   = {Phys. Rev. B},
  issue     = {16},
  volume    = {64},
  pages     = {165412},
  year      = {2001},
  month     = {10},
  publisher = {American Physical Society (APS)},
  doi       = {10.1103/physrevb.64.165412},
  url       = {https://dx.doi.org/10.1103/physrevb.64.165412}
}

@article{Hiraoka_NatCommun_8_16012_2017,
  author    = {Hiraoka, R. and Minamitani, E. and Arafune, R. and Tsukahara, N. and Watanabe, S. and Kawai, M. and Takagi, N.},
  title     = {{Single}-molecule quantum dot as a {Kondo} simulator},
  journal   = {Nat. Commun.},
  issue     = {1},
  volume    = {8},
  pages     = {16012},
  year      = {2017},
  month     = {6},
  publisher = {Springer Science and Business Media LLC},
  doi       = {10.1038/ncomms16012},
  url       = {https://dx.doi.org/10.1038/ncomms16012}
}

@article{Wahl_PhysRevLett_98_056601_2007,
  author    = {Wahl, P. and Simon, P. and Diekh{\"o}ner, L. and Stepanyuk, V. S. and Bruno, P. and Schneider, M. A. and Kern, K.},
  title     = {{Exchange} {Interaction} between {Single} {Magnetic} {Adatoms}},
  journal   = {Phys. Rev. Lett.},
  issue     = {5},
  volume    = {98},
  pages     = {056601},
  year      = {2007},
  month     = {1},
  publisher = {American Physical Society (APS)},
  doi       = {10.1103/physrevlett.98.056601},
  url       = {https://dx.doi.org/10.1103/physrevlett.98.056601}
}

@article{wan2023evidence,
  title={Evidence for ground state coherence in a two-dimensional Kondo lattice},
  author={Wan, Wen and Harsh, Rishav and Meninno, Antonella and Dreher, Paul and Sajan, Sandra and Guo, Haojie and Errea, Ion and de Juan, Fernando and Ugeda, Miguel M},
  journal={Nature communications},
  volume={14},
  number={1},
  pages={7005},
  year={2023},
  publisher={Nature Publishing Group UK London}
}

@article{kondo1964resistance,
  title={Resistance minimum in dilute magnetic alloys},
  author={Kondo, Jun},
  journal={Progress of {T}heoretical {P}hysics},
  volume={32},
  number={1},
  pages={37--49},
  year={1964},
  publisher={Oxford University Press}
}

@article{Anderson_PhysRevB_1_4464_1970,
  author    = {Anderson, P. W. and Yuval, G. and Hamann, D. R.},
  title     = {{Exact} {Results} in the {Kondo} {Problem}. {II}. {Scaling} {Theory}, {Qualitatively} {Correct} {Solution}, and {Some} {New} {Results} on {One}-{Dimensional} {Classical} {Statistical} {Models}},
  journal   = {Phys. Rev. B},
  issue     = {11},
  volume    = {1},
  pages     = {4464--4473},
  year      = {1970},
  month     = {6},
  publisher = {American Physical Society (APS)},
  doi       = {10.1103/physrevb.1.4464},
  url       = {https://dx.doi.org/10.1103/physrevb.1.4464}
}

@article{khajetoorians2019creating,
  title={Creating designer quantum states of matter atom-by-atom},
  author={Khajetoorians, Alexander A and Wegner, Daniel and Otte, Alexander F and Swart, Ingmar},
  journal={Nature Reviews Physics},
  volume={1},
  number={12},
  pages={703--715},
  year={2019},
  publisher={Nature Publishing Group UK London}
}

@article{khajetoorians2012atom,
  title={Atom-by-atom engineering and magnetometry of tailored nanomagnets},
  author={Khajetoorians, Alexander Ako and Wiebe, Jens and Chilian, Bruno and Lounis, Samir and Bl{\"u}gel, Stefan and Wiesendanger, Roland},
  journal={Nature Physics},
  volume={8},
  number={6},
  pages={497--503},
  year={2012},
  publisher={Nature Publishing Group UK London}
}

@article{Spinelli_NatMater_13_782_2014,
  author    = {Spinelli, A. and Bryant, B. and Delgado, F. and Fern{\'a}ndez-Rossier, J. and Otte, A. F.},
  title     = {{Imaging} of spin waves in atomically designed nanomagnets},
  journal   = {Nat. Mater.},
  issue     = {8},
  volume    = {13},
  pages     = {782--785},
  year      = {2014},
  month     = {7},
  publisher = {Springer Science and Business Media LLC},
  doi       = {10.1038/nmat4018},
  url       = {https://dx.doi.org/10.1038/nmat4018}
}

@article{Choi_RevModPhys_91_041001_2019,
  author    = {Choi, Deung-Jang and Lorente, Nicolas and Wiebe, Jens and von Bergmann, Kirsten and Otte, Alexander F. and Heinrich, Andreas J.},
  title     = {\textit{Colloquium}: {Atomic} spin chains on surfaces},
  journal   = {Rev. Mod. Phys.},
  issue     = {4},
  volume    = {91},
  pages     = {041001},
  year      = {2019},
  month     = {10},
  publisher = {American Physical Society (APS)},
  doi       = {10.1103/revmodphys.91.041001},
  url       = {https://dx.doi.org/10.1103/revmodphys.91.041001}
}

@article{LaShell1996_Au111_Rashba_PRL,
  author  = {LaShell, S. and McDougall, B. A. and Jensen, E.},
  title   = {Spin Splitting of an Au(111) Surface State Band Observed with Angle Resolved Photoelectron Spectroscopy},
  journal = {Phys. Rev. Lett.},
  year    = {1996},
  volume  = {77},
  pages   = {3419--3422},
  doi     = {10.1103/PhysRevLett.77.3419}
}

@article{Reinert2001_Lgap_CuAgAu_PRB,
  author  = {Reinert, F. and Nicolay, G. and Schmidt, S. and Ehm, D. and H{\"u}fner, S.},
  title   = {Direct measurements of the L-gap surface states on the (111) face of noble metals by photoelectron spectroscopy},
  journal = {Phys. Rev. B},
  year    = {2001},
  volume  = {63},
  pages   = {115415},
  doi     = {10.1103/PhysRevB.63.115415}
}

@article{Nicolay2001_AuAg111_SpinOrbit_PRB,
  author  = {Nicolay, G. and Reinert, F. and H{\"u}fner, S. and Blaha, P.},
  title   = {Spin-orbit splitting of the L-gap surface state on Au(111) and Ag(111)},
  journal = {Phys. Rev. B},
  year    = {2001},
  volume  = {65},
  pages   = {033407},
  doi     = {10.1103/PhysRevB.65.033407}
}

@article{Hoesch2004_Au111_SpinStructure_PRB,
  author  = {Hoesch, M. and Muntwiler, M. and Petrov, V. N. and Hengsberger, M. and Patthey, L. and Shi, M. and Falub, M. C. and Greber, T. and Osterwalder, J.},
  title   = {Spin structure of the Shockley surface state on the (111) surface of gold},
  journal = {Phys. Rev. B},
  year    = {2004},
  volume  = {69},
  pages   = {241401},
  doi     = {10.1103/PhysRevB.69.241401}
}

@article{Tamai2013_Cu111_Rashba_PRB,
  author  = {Tamai, A. and Meevasana, W. and King, P. D. C. and Nicholson, C. W. and de la Torre, A. and Rozbicki, E. and Baumberger, F.},
  title   = {Spin-orbit splitting of the Shockley surface state on Cu(111)},
  journal = {Phys. Rev. B},
  year    = {2013},
  volume  = {87},
  pages   = {075113},
  doi     = {10.1103/PhysRevB.87.075113}
}

@article{Hufner2008_Lgap_NobleMetals_ZPhyChem,
  author  = {H{\"u}fner, S. and Reinert, F. and Schmidt, S. and Nicolay, G. and Forster, F.},
  title   = {Photoemission Investigation of the L\={-}-Gap Surface States on Clean and Rare Gas-Covered Noble Metal (111)-Surfaces},
  journal = {Zeitschrift f{\"u}r Physikalische Chemie},
  year    = {2008},
  volume  = {222},
  number  = {2-3},
  pages   = {407--431},
  doi     = {10.1524/zpch.2008.222.2-3.407}
}

@article{Ast2007_BiAg111_GiantRashba_PRL,
  author  = {Ast, C. R. and Henk, J. and Ernst, A. and Moreschini, L. and Falub, M. C. and Pacil{\'e}, D. and Bruno, P. and Kern, K. and Grioni, M.},
  title   = {Giant Spin Splitting through Surface Alloying},
  journal = {Phys. Rev. Lett.},
  year    = {2007},
  volume  = {98},
  pages   = {186807},
  doi     = {10.1103/PhysRevLett.98.186807}
}

@article{Gierz2010_StructuralInfluence_Rashba_ArXiv,
  author  = {Gierz, I. and Stadtm{\"u}ller, B. and Vuorinen, J. and Lindroos, M. and Meier, F. and Dil, J. H. and Kern, K. and Ast, C. R.},
  title   = {The Structural Influence on the Rashba-type Spin-Splitting in Surface Alloys},
  journal = {arXiv:1003.2351},
  year    = {2010},
  url     = {https://arxiv.org/abs/1003.2351}
}

@article{Meier2009_BiPbAg_TunableFermi_PRB,
  author  = {Meier, F. and Petrov, V. and Guerrero, S. and Mudry, C. and Patthey, L. and Osterwalder, J. and Dil, J. H.},
  title   = {Unconventional Fermi surface spin textures in the Bi\_xPb\_{1-x}/Ag(111) surface alloy},
  journal = {Phys. Rev. B},
  year    = {2009},
  volume  = {79},
  pages   = {241408},
  doi     = {10.1103/PhysRevB.79.241408}
}

@article{Pacile2006_PbAg111_PRB,
  author  = {Pacil{\'e}, D. and Ast, C. R. and Papagno, M. and Da Silva, A. J. R. and Mazzarello, R. and {\'O}rdej{\'o}n, P. and Carbone, C. and Grioni, M.},
  title   = {Electronic structure of an ordered Pb/Ag(111) surface alloy: Theory and experiment},
  journal = {Phys. Rev. B},
  year    = {2006},
  volume  = {73},
  pages   = {245429},
  doi     = {10.1103/PhysRevB.73.245429}
}

@article{Varykhalov2012_Ir111_GiantRashba_PRL,
  author  = {Varykhalov, A. and Marchenko, D. and Scholz, M. R. and Rienks, E. D. L. and Kim, T. K. and Bihlmayer, G. and S{\'a}nchez-Barriga, J. and Rader, O.},
  title   = {Ir(111) Surface State with Giant Rashba Splitting Persists under Graphene in Air},
  journal = {Phys. Rev. Lett.},
  year    = {2012},
  volume  = {108},
  pages   = {066804},
  doi     = {10.1103/PhysRevLett.108.066804}
}

@book{Weinberg_QFT1,
  author    = {Steven Weinberg},
  title     = {The Quantum Theory of Fields. Vol. I: Foundations},
  publisher = {Cambridge University Press},
  year      = {1995},
  isbn      = {9780521550017},
  doi       = {10.1017/CBO9781139644167}
}

@book{Plemelj_RiemannKlein_1964,
  author     = {Josip Plemelj},
  title      = {Problems in the Sense of Riemann and Klein},
  translator = {J. R. M. Radok},
  series     = {Interscience Tracts in Pure and Applied Mathematics},
  number     = {16},
  publisher  = {Interscience Publishers (John Wiley \& Sons)},
  address    = {New York},
  year       = {1964}
}

@article{OsorioNonLocal,
  title = {Nonlocal damping of spin waves in a magnetic insulator induced by normal, heavy, or altermagnetic metallic overlayer: A Schwinger-Keldysh field theory approach},
  author = {Reyes-Osorio, Felipe and Nikoli\ifmmode \acute{c}\else \'{c}\fi{}, Branislav K.},
  journal = {Phys. Rev. B},
  volume = {110},
  issue = {21},
  pages = {214432},
  numpages = {9},
  year = {2024},
  month = {Dec},
  publisher = {American Physical Society},
  doi = {10.1103/PhysRevB.110.214432},
  url = {https://link.aps.org/doi/10.1103/PhysRevB.110.214432}
}

@article{PhysRevX.5.021025,
  title = {Nonreciprocal Photon Transmission and Amplification via Reservoir Engineering},
  author = {Metelmann, A. and Clerk, A. A.},
  journal = {Phys. Rev. X},
  volume = {5},
  issue = {2},
  pages = {021025},
  numpages = {16},
  year = {2015},
  month = {Jun},
  publisher = {American Physical Society},
  doi = {10.1103/PhysRevX.5.021025},
  url = {https://link.aps.org/doi/10.1103/PhysRevX.5.021025}
}

@article{Madhavan1998ScienceKondo,
  author  = {Madhavan, V. and Chen, W. and Jamneala, T. and Crommie, M. F. and Wingreen, N. S.},
  title   = {Tunneling into a Single Magnetic Atom: Spectroscopic Evidence of the Kondo Resonance},
  journal = {Science},
  year    = {1998},
  volume  = {280},
  number  = {5363},
  pages   = {567--569},
  doi     = {10.1126/science.280.5363.567}
}

@article{Knorr2002PRL096804,
  author  = {Knorr, N. and Schneider, M. A. and Diekh{\"o}ner, L. and Wahl, P. and Kern, K.},
  title   = {Kondo Effect of Single Co Adatoms on Cu Surfaces},
  journal = {Physical Review Letters},
  year    = {2002},
  volume  = {88},
  number  = {9},
  pages   = {096804},
  doi     = {10.1103/PhysRevLett.88.096804}
}

@article{Jamneala2000PRB9990,
  author  = {Jamneala, T. and Madhavan, V. and Chen, W. and Crommie, M. F.},
  title   = {Scanning tunneling spectroscopy of transition-metal impurities at the surface of gold},
  journal = {Physical Review B},
  year    = {2000},
  volume  = {61},
  number  = {15},
  pages   = {9990--9993},
  doi     = {10.1103/PhysRevB.61.9990}
}

@article{Limot2005PRL036805,
  author  = {Limot, L. and Pehlke, E. and Kr{\"o}ger, J. and Berndt, R.},
  title   = {Surface-state localization at adatoms},
  journal = {Physical Review Letters},
  year    = {2005},
  volume  = {94},
  number  = {3},
  pages   = {036805},
  doi     = {10.1103/PhysRevLett.94.036805}
}

@article{PhysRev.120.91,
  title = {Anisotropic Superexchange Interaction and Weak Ferromagnetism},
  author = {Moriya, T\^oru},
  journal = {Phys. Rev.},
  volume = {120},
  issue = {1},
  pages = {91--98},
  numpages = {0},
  year = {1960},
  month = {Oct},
  publisher = {American Physical Society},
  doi = {10.1103/PhysRev.120.91},
  url = {https://link.aps.org/doi/10.1103/PhysRev.120.91}
}

@article{DZYALOSHINSKY1958241,
title = {A thermodynamic theory of “weak” ferromagnetism of antiferromagnetics},
journal = {Journal of Physics and Chemistry of Solids},
volume = {4},
number = {4},
pages = {241-255},
year = {1958},
issn = {0022-3697},
doi = {https://doi.org/10.1016/0022-3697(58)90076-3},
url = {https://www.sciencedirect.com/science/article/pii/0022369758900763},
author = {I. Dzyaloshinsky}
}
\end{document}